\documentclass[%
 reprint,
superscriptaddress,
amsmath,amssymb,
aps,
floatfix,
]{revtex4-2}

\usepackage{graphicx}% Include figure files
\usepackage{dcolumn}% Align table columns on decimal point
\usepackage[utf8]{inputenc}
\DeclareUnicodeCharacter{2212}{\textminus}
\usepackage{bm}% bold math
\usepackage[colorlinks]{hyperref}% add hypertext capabilities
\usepackage{float}
\usepackage{makecell}
\usepackage{comment}
\usepackage[thinlines]{easytable}
\usepackage{hhline}
\usepackage{braket, siunitx, placeins}
\usepackage{epsfig}
\usepackage{siunitx}

\begin{document}

\preprint{APS/123-QED}

\title{$ZZ$-Free Two-Transmon CZ Gate Mediated by a Fluxonium Coupler}% Force line breaks with \\

\def\RLEaffil{Research Laboratory of Electronics, Massachusetts Institute of Technology, Cambridge, MA 02139, USA}
\def\LLaffil{Lincoln Laboratory, Massachusetts Institute of Technology, Lexington, MA 02421, USA}
\def\Physaffil{Department of Physics, Massachusetts Institute of Technology, Cambridge, MA 02139, USA}
\def\EECSaffil{Department of Electrical Engineering and Computer Science, Massachusetts Institute of Technology, Cambridge, MA 02139, USA}
\def\affilAQ{Google Quantum AI, Cambridge, MA 02139}
\def\affilQM{Quantum Machines, Tel Aviv-Yafo, 6721407, Israel}
\def\affilIBM{IBM Quantum, Cambridge, MA 02139}
\def\affilTriarii{Triarii Research, Netanya, Israel}

\newcommand{\fref}[1]{Fig.~\ref{#1}}
\newcommand{\eref}[1]{Eq.~(\ref{#1})}
\newcommand{\aref}[1]{Appendix~\ref{#1}}

\author{Junyoung~An}
\affiliation{\RLEaffil}
\affiliation{\EECSaffil}

\author{Helin~Zhang}
\affiliation{\RLEaffil}

\author{Qi~Ding}
\affiliation{\RLEaffil}
\affiliation{\EECSaffil}

\author{Leon~Ding}
\altaffiliation[Present address: ]{\affilAQ}
\affiliation{\RLEaffil}
\affiliation{\Physaffil}

\author{Youngkyu~Sung}~
\altaffiliation[Present address: ]{\affilAQ}
\affiliation{\RLEaffil}
\affiliation{\EECSaffil}

\author{Roni~Winik}
\altaffiliation[Present address: ]{\affilTriarii}
\affiliation{\RLEaffil}

\author{Junghyun~Kim}
\affiliation{\RLEaffil}
\affiliation{\EECSaffil}

\author{Ilan~T.~Rosen}
\altaffiliation[Present address: ]{\affilIBM}
\affiliation{\RLEaffil}

\author{Kate~Azar}
\affiliation{\RLEaffil}
\affiliation{\EECSaffil}
% \affiliation{\LLaffil}

\author{Renee~DePencier~Pi\~nero}
\affiliation{\LLaffil}

\author{Jeffrey~M.~Gertler}
\affiliation{\LLaffil}

\author{Michael~Gingras}
\affiliation{\LLaffil}

% \author{Thomas~M.~Hazard}
% \affiliation{\LLaffil}

\author{Bethany~M.~Niedzielski}
\affiliation{\LLaffil}

\author{Hannah~Stickler}
\affiliation{\LLaffil}

\author{Mollie~E.~Schwartz}
\affiliation{\LLaffil}

\author{Joel~\^I-j.~Wang}
\affiliation{\RLEaffil}

\author{Terry~P.~Orlando}
\affiliation{\RLEaffil}
\affiliation{\EECSaffil}

\author{Simon~Gustavsson}
\altaffiliation[Present address: ]{\affilAQ}
\affiliation{\RLEaffil}

\author{Max~Hays}
\affiliation{\RLEaffil}

\author{Jeffrey~A.~Grover}
\affiliation{\RLEaffil}

\author{Kyle~Serniak}
\affiliation{\RLEaffil}
\affiliation{\LLaffil}

\author{William~D.~Oliver}
\affiliation{\RLEaffil}
\affiliation{\EECSaffil}
\affiliation{\Physaffil}

\email{william.oliver@mit.edu}

\date{\today}% It is always \today, today,

\begin{abstract}
Eliminating residual $ZZ$ interactions in a two-qubit system is essential for reducing coherent errors during quantum operations. In a superconducting circuit platform, coupling two transmon qubits via a transmon coupler has been shown to effectively suppress residual $ZZ$ interactions. However, in such systems, perfect cancellation usually requires the qubit-qubit detuning to be smaller than the individual qubit anharmonicities, which exacerbates frequency crowding and microwave crosstalk. To address this limitation, we introduce TFT (Transmon-Fluxonium-Transmon) architecture, wherein two transmon qubits are coupled via a fluxonium qubit. The coupling mediated by the fluxonium eliminates residual $ZZ$ interactions even for transmons detuned larger than their anharmonicities. We experimentally identified zero-$ZZ$ interaction points at qubit-qubit detunings of \SI{409}{MHz} and \SI{616}{MHz} from two distinct TFT devices. We then implemented an adiabatic, coupler-flux-biased controlled-Z gate on both devices, achieving CZ gate fidelities of {99.64(6)\%} and {99.68(8)\%}.
\end{abstract}
\maketitle

%\tableofcontents

\section{\label{sec:introduction} INTRODUCTION}

Transmon-based quantum circuits are a leading hardware platform in the field of quantum computing. Single-qubit gates with an error rate below $10^{-4}$ \cite{ZhiyuanLi2023_1QBtransmon9999, Bland2025_1QBtransmon9999} and two-qubit gates with an error rate below or near $10^{-3}$ ~\cite{Marxer2025_IQM999fidelity, Glaser2024_TTT9992QB} have been demonstrated in transmon-based quantum processors. In addition, transmon qubits \cite{Koch2007_Transmon} support fast, high-fidelity reset \cite{Zhou2021_FastResetTransmon} and readout \cite{Spring2025_intrinsicPCfilterRO} of their computational states. The properties of transmon qubits and couplers have made them a favored option in quantum processors capable of detecting and correcting errors in quantum circuits.~\cite{Acharya2023_Google_d=5surfacecode, Willow2025, Wang2025_qLDPC, lacroix2024_scalinglogiccolorcode}.

One of the remaining major challenges of an all-transmon system is the unwanted residual $ZZ$ interaction ($\zeta_{ZZ}$) between two qubits. Residual $ZZ$ interactions are a major source of coherent error during simultaneous single-qubit~\cite{Gambetta2012_Sim1QBRBandZZ, Mundada2019_ZZxtalkCancel} and two-qubit gates~\cite{Zhao2023_CRgateZZmitigation, Sheldon2016_CRgateZZ}. Hence, canceling the residual $ZZ$ interactions is a crucial step in further improving operational fidelity.  

Residual $ZZ$ interactions are unavoidable with capacitively coupled two transmon qubits, due to the asymmetric interactions between each computational state and the non-computational states~\cite{JaseungKu2020_UnwantedZZ}. Furthermore, while increasing the gate speed serves to reduce the error rate due to decoherence, it often requires larger coupling strengths, which increases the residual $ZZ$ interaction and thereby counteracts the benefit. One way to overcome this trade-off is to add a transmon coupler, which effectively reduces \cite{YuanXu2020_TTTCZgate, Collodo2020_TTT} or even eliminates \cite{Sung2021_TTT, Marxer2023_TTTCZgate} the residual $ZZ$ interaction at the cost of added complexity. 

Although effective, the transmon coupler is not a complete solution to residual $ZZ$ interactions. In most implementations, the data qubit must be operated in the straddling regime---where the qubit-qubit detuning is smaller than the qubit anharmonicities---to completely cancel the $ZZ$ interactions \cite{Sung2021_TTT, Stehlik2021_TTT}. When operated in this regime, the qubits are vulnerable to microwave crosstalk and frequency crowding owing to the relatively small anharmonicity of a transmon \cite{Zhang2022_freqcollision, Hertzberg2021_freqcollision}. Hence, alternative coupling strategies have been proposed \cite{Goto2022_DTStheory, Heunisch2023_CSFQcplrtheory, JiangLing2025_ZZzeroTTT} to completely cancel $ZZ$ interactions while operating the transmon qubits outside of the straddling regime. 

Here, we introduce the TFT (Transmon - Fluxonium - Transmon) architecture, which uses a fluxonium as a coupling element between two transmons to enable $ZZ$ cancellation outside the straddling regime. We experimentally verified that a TFT circuit could cancel static $ZZ$ interactions for qubit-qubit detuning exceeding transmon qubit anharmonicities. In addition, flux-biasing the fluxonium coupler can increase the $ZZ$ interaction, thereby allowing the implementation of a coupler-flux-biased CZ gate. We measured CZ gate fidelities of 99.64(6)\% and 99.68(8)\% for two separate TFT devices using interleaved randomized benchmarking. Lastly, we discuss the advantages and challenges in using the TFT system as a primitive for larger-scale processors.

\section{\label{sec:principle_of_tft} TFT PRINCIPLE OF OPERATION}

\begin{figure}[t]
    \centering
    \includegraphics[width=0.48\textwidth]{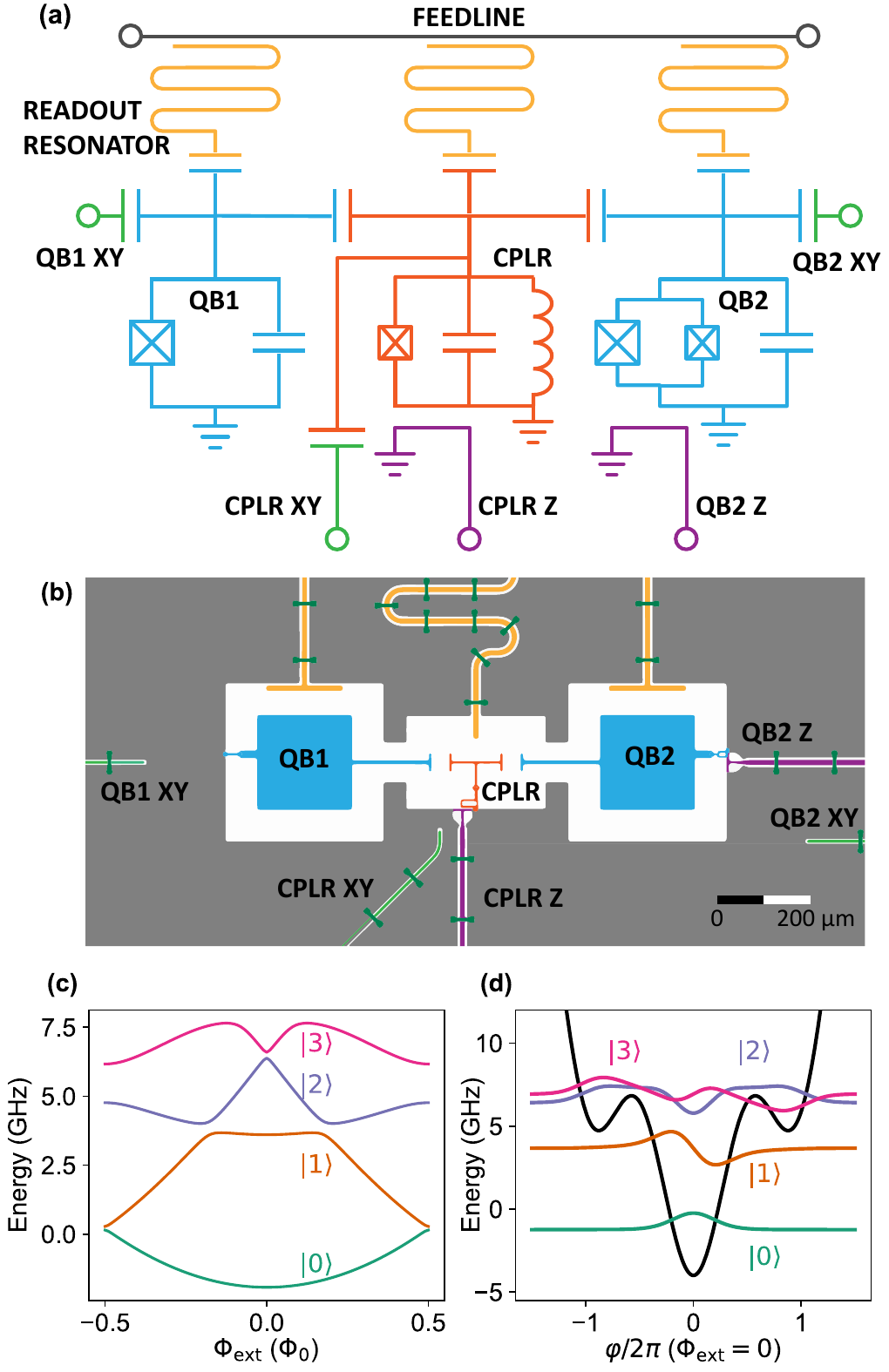}
    \caption{(a) Schematic of the TFT circuit. The color of each element matches the color on the device image (b). (b) False-colored gds design image of the TFT device.  (c) Typical energy levels of the fluxonium coupler. (d) Potential energy (black line) and the phase-space wavefunctions (colored lines) of the first four energy eigenstates of the fluxonium coupler at $\Phi_{\mathrm{ext}}=0$. The energy parameters used to create (c), (d) are $E_{C}/h=1.0$, $E_{J}/h=5.0$, $E_{L}/h=0.5$ in units of GHz.}
    \label{fig:tft_schematics}
\end{figure}

A TFT system comprises two transmons coupled via a fluxonium qubit with capacitances as shown in \fref{fig:tft_schematics}(a)-(b). The ground- and first-excited states of each transmon form the single-qubit logical states. The fluxonium coupler energy levels serve to mediate an effective interaction between the two transmon qubits. 

Fluxonium is a superconducting qubit comprising a single small Josephson junction shunted by an inductor, typically realized using an array of larger Josephson junctions. The shunt inductance $L$ of the fluxonium should be sufficiently large such that the inductive energy $E_{L}=\Phi_{0}^{2}/2L$ is much smaller than the small-junction Josephson junction energy $E_{J}=\Phi_{0} I_{c}/2\pi$ \cite{Manucharyan2009_fluxonium, LongNguyen2019_FluxoniumReview}, where $\Phi_{0} = h/2e$ is the superconducting magnetic flux quantum, and $I_{c}$ is the critical current of the small junction. In this $E_{L} \ll E_{J} $ regime, the fluxonium exhibits a double-well potential when flux biased at $\Phi_{0}/2$, similar to a persistent current flux qubit \cite{Orlando1999_FluxQubit}. The symmetric, anti-symmetric superposition of the two persistent-current states form the first two energy eigenstates, with its energy splitting determined by the tunneling amplitude between two potential wells. This tunneling amplitude can be suppressed as an exponential function of the ratio between $E_{J}$ and the fluxonium charging energy $E_{C}=e^{2}/2C$, where $C$ is the total shunting capacitance across the small junction. The suppression of tunneling amplitude leads to near-degeneracy and a reduced energy decay rate of the first-two energy eigenstates.

When used as a coupler, however, we bias the fluxonium near zero flux, where it exhibits the following features. First, the wavefunctions of the ground and first-excited states are localized within a single potential well, forming an intra-well (plasmon-like) transition. In contrast, the second and third excited states are extended over several potential wells and weakly hybridize to form a nearly degenerate level pairs [\fref{fig:tft_schematics}(c)-(d)]. These conditions are achieved for $E_{J}/E_{L}\approx$ 7--10 with $E_{C}/h \approx$ \SI{1}{GHz}; the specific values will depend on the transmon qubit frequencies and the qubit-coupler interaction strength. In such a fluxonium coupler, adjusting the coupler flux continuously tunes the $ZZ$ interaction between positive and negative values, thereby guaranteeing a zero-$ZZ$ interaction bias point in-between.

\begin{figure}
    \centering
    \includegraphics[width=0.48\textwidth]{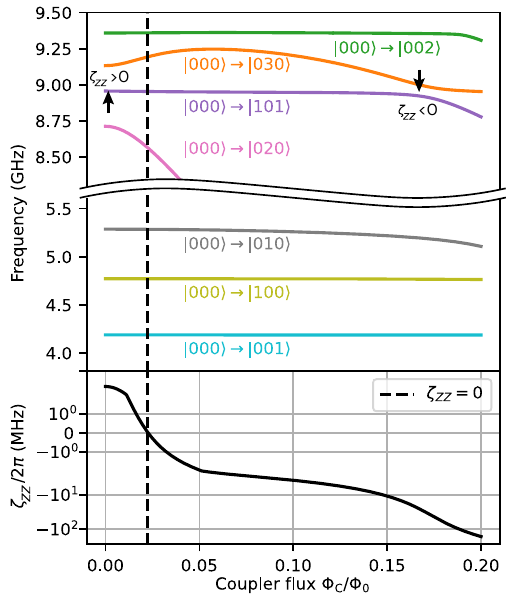}
    \caption{Transitions and $ZZ$ interactions of a model TFT system as a function of coupler flux. The coupler parameters are set to $E_{C}/h=\SI{0.9}{GHz}$, $E_{J}/h=\SI{4.8}{GHz}$, $E_{L}/h=\SI{0.55}{GHz}$. The two transmon qubit parameters are set as $\omega_{\textrm{Q1}}/2\pi=\SI{4.8}{GHz}$, $\omega_{\textrm{Q2}}/2\pi=\SI{4.2}{GHz}$, and $E_{C1}/h$=$E_{C2}/h=\SI{0.2}{GHz}$. The coupling strength are $J_{1c}/h=J_{2c}/h=\SI{0.15}{GHz}$ and $J_{12}/h=\SI{0.015}{GHz}$. The black dashed vertical line indicates the coupler flux where the $ZZ$ interaction becomes zero. The two black arrows indicate the flux biases where the $|101 \rangle$ state is effectively being ``pushed up" (positive $\zeta_{ZZ}$) or ``pushed down" (negative $\zeta_{ZZ}$). The states are labeled assuming adiabatic transition from zero flux bias.}
    \label{fig:transition_zz_vs_cplrflux}
\end{figure}

\subsection{Cancelling and Enhancing the $ZZ$ Interaction} 

To illustrate how the fluxonium coupler controls the $ZZ$ interaction in the system, we show the energy transitions from the ground state to higher excited states after diagonalizing the Hamiltonian of the TFT system:
\begin{align} \label{eqn:tft_hamiltonian}
\begin{split}
    \hat{H} = & \sum_{i=1,2} \left( 4 E_{C,i} \hat{n}_{i}^{2} - E_{J,i} \cos \hat{\varphi}_{i} \right) \\ 
    & + 4 E_{C,c} \hat{n}_{c}^{2} + \frac{1}{2} E_{L,c} \hat{\varphi}_{c}^{2} - E_{J,c} \cos (\hat{\varphi}_{c} - \phi_{\textrm{ext}} ) \\
    & + J_{1c} \hat{n}_{1} \hat{n}_{c} + J_{2c} \hat{n}_{c} \hat{n}_{2} + J_{12} \hat{n}_{1} \hat{n}_{2}
\end{split}
\end{align}
where $J_{1c}, J_{2c}, J_{12}$ are the interaction energies between qubit 1 and the coupler, qubit 2 and the coupler, and the direct qubit 1 - qubit 2 interaction, respectively. The derivation of \eref{eqn:tft_hamiltonian} is given in \aref{app:tft_circuit_hamiltonian}. From the energy transitions, we calculate the $ZZ$ interaction:
\begin{align} \label{eqn:zz_definition_cplr}
    \zeta_{ZZ} = (E_{101} - E_{100} - E_{001} + E_{000})/\hbar,
\end{align}
where the subscripts indicate the excitation number of qubit 1, coupler, and qubit 2, respectively. This convention is used throughout the paper, including for the numbers inside the bra-ket notations.

The $ZZ$ interaction in the TFT system is dominated by contributions from interactions involving $|101\rangle$, $|020 \rangle$, and $|030 \rangle$. For the purpose of explanation, we assume the case $E_{020} < E_{101} < E_{030}$ at zero flux, although this is not a requirement. At zero flux, the $ZZ$ interaction is positive, dominated by the interaction between states $|101\rangle$ and $|020\rangle$. The $|101\rangle$ state is effectively ``pushed up" by $|020\rangle$, mediated by $|011\rangle$ or $|110\rangle$ since there is no direct interaction between $|101\rangle$ and $|020\rangle$ [\fref{fig:transition_zz_vs_cplrflux}]. If we start to increase the coupler flux bias from zero, the interaction with $|020\rangle$ vanishes rapidly as $|020 \rangle$ moves away from $|101\rangle$. The $ZZ$ interaction becomes zero when the positive interaction with $|020 \rangle$ cancels the negative interactions from the direct qubit-qubit interaction, mediated by $|101 \rangle$, $|200\rangle$, and $|002\rangle$. Further away from the zero-$ZZ$ point, the level repulsion between $|101\rangle$ and $|030\rangle$ starts to dominate as they become closer in frequency, resulting in a negative contribution to the overall $ZZ$ interaction as $|101\rangle$  is being ``pushed down" in frequency by $|030\rangle$. In the vicinity of the avoided crossing between $|101 \rangle$ and $|030 \rangle$, the magnitude of the $ZZ$ interaction can reach several tens of MHz, allowing a conditional phase to be accrued in a relatively short amount of time. A quantitative analysis of the $ZZ$ interaction based on a perturbative expansion is introduced in \aref{app:zz_analysis}.

Utilizing this tunable $ZZ$ interaction, a CZ gate can be implemented by adiabatically sweeping the coupler flux from the zero-$ZZ$ interaction point to the region with larger $ZZ$ interaction strength. In the next section, we present experimental validations of these concepts.

\begin{figure*}[t]
    \centering
    \includegraphics[width=0.98 \textwidth]{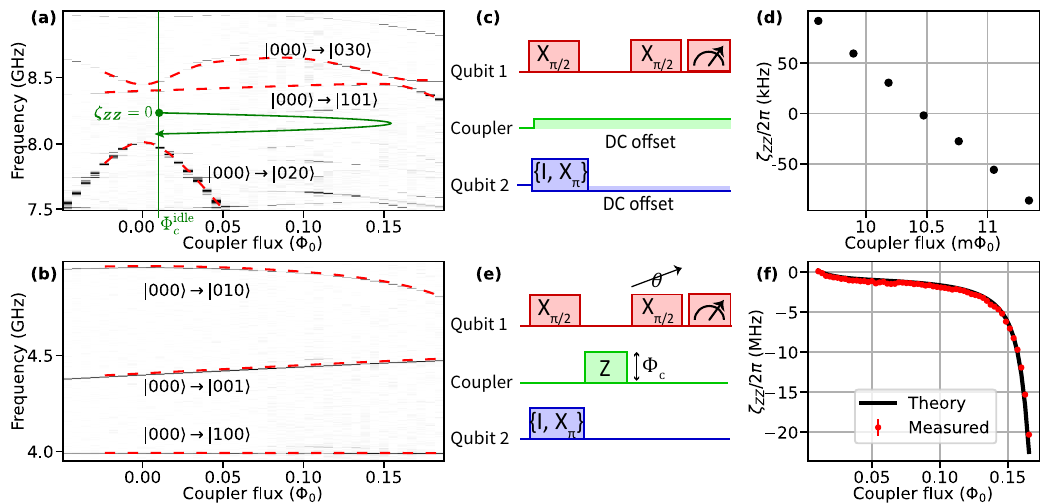}
    \caption{(a, b) Two-tone spectroscopy measurement and its fit (red dashed line) of the TFT device A. The coupler flux was swept from {-0.05}{$\Phi_{0}$} to {0.17}{$\Phi_{0}$}, which covers the operating range of the CZ gate. Spectrum (a) shows the transitions to multi-photon excitation states, including $|020 \rangle$, $|101 \rangle$, and $|030 \rangle$. The green curved arrow shows the example of CZ gate trajectory, which starts from zero-$ZZ$ flux bias (idling point). Spectrum (b) shows the single-photon excitation states, including transitions to $|100 \rangle$, $|001 \rangle$, and $|010 \rangle$. (c) Pulse sequence for the $ZZ$ interaction measurement near the idling point. When we were sweeping the coupler flux, we applied a compensating flux to qubit 2 to keep its frequency constant. (d) $ZZ$ interaction measured near the zero-$ZZ$ flux bias. We verified the sign change of the $ZZ$ interaction, and we measured $\zeta_{ZZ}/2\pi$=0.07$\pm$\SI{0.55}{kHz} at the coupler flux bias $\Phi_{c} = 0.0104 \Phi_{0}$. (e) Pulse sequence for the $ZZ$ interaction measurement away from the idling point. We apply a square-shaped flux pulse to the coupler between the two Ramsey pulses, with a varying phase $\theta$ applied to the second pi-half pulse (depicted as the straight arrow through  $\theta$) to measure the accrued conditional phase. (f) $ZZ$ interaction measurement results (red circles) that away from the idling point with a comparison to theory (black solid line).}
    \label{fig:zz_vs_cplrflux}
\end{figure*}
\begin{center}
\begin{table}[t]
    \centering
    \begin{tabular}{>{\centering}m{0.2cm} >{\centering}m{1.3cm} >{\centering}m{1.85cm} >{\centering}m{1.1cm} >{\centering}m{1.1cm} >{\centering\arraybackslash}m{1.85cm}}
        \hline
        \hline
        \multicolumn{2}{c}{Device} & $E_{J}/h$ & $E_{C}/h$ & $E_{L}/h$ & $\omega_{01}/2\pi $ \\
         \hline
         A & Qubit 1 & 9.875 & 0.225 & -  & 3.991 \\
         A & Qubit 2 & [3.897, 13.51] & 0.227 & - & [3.952, 5.369] \\
         A & Coupler & 4.372 & 0.890 & 0.511 & 4.959 \\ 
         B & Qubit 1 & 12.6 & 0.20 & - & 4.243 \\
         B & {Qubit 2} & [2.7, 9.35] & 0.187 & - & [2.97, 4.06]  \\
         B & {Coupler}  & 3.93 & 0.83 & 0.50 & 4.66  \\
        \hline
         & & $J_{1c}/h$ & $J_{2c}/h$ & $J_{12}/h$ & \\
         \hline
        A & Interaction & 0.122 & 0.134 & 0.0121 & \\
        B & Interaction & 0.228& 0.228& 0.022& \\
         \hline
         \hline
    \end{tabular}
    \caption{Energy parameters and qubit frequencies of the two TFT  devices, denoted A and B. The parameters are extracted from the spectrum as matched to the Hamiltonian model \eref{eqn:tft_hamiltonian}. The values inside the square brackets indicate the minimum and maximum values, which are relevant to the qubit 2, whose frequencies can be tuned by flux biasing. The 0-1 frequencies ($\omega_{01}/2\pi$) of the couplers are measured from their zero-$ZZ$ flux bias points. All units are in GHz.}
    \label{tab:tft_param}
\end{table}
\end{center}

\section{\label{sec:experiment_result} Basic Characterization}

\subsection{Device Geometry and Parameters}

We designed and fabricated a TFT circuit on a high-resistivity silicon wafer. Qubit~1 is a fixed-frequency transmon, and qubit 2 is a flux-tunable transmon, although the flux tunability of qubit 2 is not needed here for the CZ gate operation. Finally, the fluxonium coupler consists of a small Josephson junction shunted by a junction array, which serves as an inductor.

We measured the qubit spectra of two devices labeled A, B and determined the constituent component parameters by numerically fitting the spectra to the model Hamiltonian \eref{eqn:tft_hamiltonian}. The two devices differ in terms of their energy parameters and interaction strengths. The qubit spectrum of device A is shown in \fref{fig:zz_vs_cplrflux}(a,b), and the resulting qubit parameters are summarized in Table~\ref{tab:tft_param}. Before proceeding to the $ZZ$ interaction measurement and gate calibration, we intentionally tuned qubit 2 to set the system outside the straddling regime. For device A, the frequency of qubit 1 was fixed at \SI{3.991}GHz and the frequency of qubit 2 was flux-biased to \SI{4.40}{GHz} ($\Delta_{12}/2\pi=$\SI{409}{MHz}). 
For device B, qubit 1 was fixed at \SI{4.243}{GHz} and qubit 2 was flux-biased to \SI{3.627}{GHz} ($\Delta_{12}/2\pi = 616 \textrm{ MHz}$). 

The rest of the main text will focus on the measurement results from device A. The results from device B are summarized in Appendix~\ref{app:tft1_summary}.

\subsection{$ZZ$ Interaction Measurement}

After setting the qubit-qubit detuning, we measured the $ZZ$ interaction of the TFT system as a function of coupler flux. The measurement was done in two ranges: a fine sweep near zero-$ZZ$ point and a wider sweep. First, to measure the $ZZ$ interaction near its null point, we characterized the qubit 1 frequency using a Ramsey sequence [\fref{fig:zz_vs_cplrflux}(c)], once with qubit 2 in its ground state and then again qubit 2 in its excited state. The difference in qubit 1 frequencies conditioned on the logical state of qubit 2 is the $ZZ$ interaction. As we swept the coupler flux, $ZZ$ interaction changed sign from positive to negative, confirming the existence of zero-$ZZ$ point near the coupler flux bias $\Phi_{\mathrm{c}}^{\textrm{idle}} = 0.0104 {\Phi_{0}}$ [\fref{fig:zz_vs_cplrflux}(d)]. We hereafter call this the ``idling point.'' At the idling point, we measured a $ZZ$ interaction of 0.07$\pm$\SI{0.55}{kHz}.

Next, we performed a wider sweep of the $ZZ$ interaction moving from the idling point toward the avoided crossing between $|101\rangle$ and $|030 \rangle$. In this measurement, we applied a square-shaped flux pulse to the coupler during the free evolution of qubit 1, as shown in \fref{fig:zz_vs_cplrflux}(e). We then measured qubit 1 population as a function of the phase $\theta$ of the second $\pi/2$ pulse. Depending on the state of the qubit 2, the resulting signals exhibited a relative phase shift of $\Delta \phi =\zeta_{ZZ} \tau $ where $\tau$ is the duration of the coupler flux pulse. We observed a good agreement between the theory and the measurement results [\fref{fig:zz_vs_cplrflux}(f)]. 

We also highlight that we could eliminate the $ZZ$ interaction when the two qubits are in the straddling regime. We brought the frequency of qubit 2 close to that of qubit 1, and confirmed the change of sign in the $ZZ$ interaction near its expected idling point. The results are presented in Appendix~\ref{app:TFT_$ZZ$_straddling regime}, \fref{fig:zz_at_straddling_regime}.

\begin{figure*}[t]
    \centering
    \includegraphics[width=0.98\textwidth]{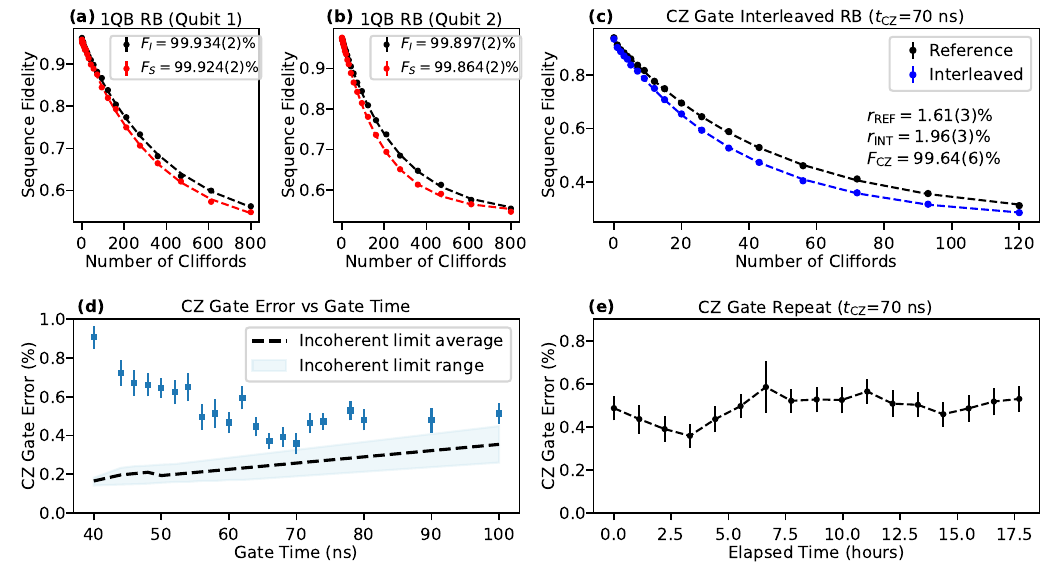}
    \caption{(a, b) Single-qubit individual (black) and simultaneous (red) randomized benchmarking (RB) results of qubit 1 (a) and qubit 2 (b). The sequence fidelities were calculated by averaging the results from 30 random Clifford sequences. (c) Two-qubit Clifford RB at a \SI{70}{ns} CZ gate time ($t_{\mathrm{CZ}}$). We obtained an average CZ gate fidelity of 99.64(6)\% from 40 random Clifford sequences. (d) CZ gate error measurements gate time ranging from \SI{40}{ns} to \SI{100}{ns}. We observed an optimum gate time near $t_{\mathrm{CZ}}=$\SI{70}{ns}. The average incoherent error limit (black dashed line) and its uncertainty (light blue region) is determined by the coherence time measurement results of the two qubits, which are detailed in \aref{app:qubit_coherence_repeat} and Table \ref{tab:tft_coherence_time_app}. (e) CZ gate fidelity measurement for a \SI{70}{ns} gate time repeated over 17 hours. During the CZ gate randomized benchmarking (c - e), the coupler and qubit flux biases were recalibrated every 30 minutes to compensate for flux drifts. The error bars on each point of the RB (a - c), mostly obscured by markers, indicate the standard error of the mean. The error bars of the average gate fidelities in (d, e) are obtained from the uncertainty of the curve fit followed by error propagation.}
    \label{fig:cz_rb}
\end{figure*}

\section{Gate Calibration and Benchmarking}

\subsection{Single-Qubit Gate Benchmarking}

We calibrated single-qubit gates on device A at the idling point following ~\cite{LeonDing2023_FTF}. The single-qubit gates were implemented with a \SI{40}{ns} microwave drive with a cosine-shaped pulse envelope. The gates were randomly selected by 24 single-qubit Clifford gates, generated by the set of microwave gates \{$I, \pm X_{\pi}, \pm Y_{\pi},\pm X_{\pi/2}, \pm Y_{\pi/2}$\}. The average single-qubit gate fidelities were measured via Clifford randomized benchmarking \cite{Knill2008_CliffordRB, Magesan2012_CliffordRB}. For qubit 1, we obtained average gate fidelities of {99.934(2)\%} for individual randomized benchmarking and {99.924(2)\%} for simultaneous randomized benchmarking [\fref{fig:cz_rb}(a)]. For qubit 2, we measured average gate fidelities of {99.897(2)\%} and {99.864(2)\%} for individual and simultaneous randomized benchmarking, respectively [\fref{fig:cz_rb}(b)]. The fidelities of individual and simultaneous measurements were nearly identical for both qubits, which is consistent with $ZZ$ crosstalk being suppressed.

\subsection{CZ Gate Implementation and Calibration}

In device A, the CZ gate was implemented using a flux pulse based on a Slepian shape, which is designed to suppress leakage during the adiabatic gate operation~\cite{Martinis2014_Slepian, QiDingEQuS2025}. However, the transition frequency from $|101 \rangle$ to $|030 \rangle$ is nonmonotonic from the idling point to the avoided crossing, which complicates the pulse construction. 

To address this issue, we divided the time interval and constructed the total flux pulse by connecting two different pulses. First, for the time interval $t < t_{\mathrm{rise}}$, we built a pulse whose value increases at a polynomial rate in time ($\Phi_{\textrm{ext}}(t) = At^{\alpha}+\Phi_{\textrm{ext}}^{0}$). In device A, we empirically found the polynomial shape as a simple model to optimize the pulse shape to suppress leakage to $|030 \rangle$ during $t< t_{\mathrm{rise}}$. The coefficient $A$ is determined such that at time $t=t_{\mathrm{rise}}$, the pulse reaches the point where the frequency difference between $|101 \rangle$ and $|030 \rangle$ becomes maximum. We chose $\alpha$, $t_{\mathrm{rise}}$ to reduce leakage during $t < t_{\mathrm{rise}}$ based on the numerical simulation in  \aref{app:flux_pulse_optimization}. For $ t \geq t_{\mathrm{rise}}$, we built a Slepian-shaped flux pulse based on the frequencies between two energy levels, $|101 \rangle$ and $|030 \rangle$, and their effective interaction strength. This interaction strength is obtained from the transition frequency at their avoided crossing [\fref{fig:transition_zz_vs_cplrflux}(a)]. The detailed procedure for the gate implementation is in Appendix~\ref{app:flux_pulse_optimization}.

Next, we calibrated the CZ gate with three control variables: rise time ($t_{\mathrm{rise}}$), Slepian final frequency, and Slepian bandwidth. We found a set of these three variables that yields the minimum leakage from $|101 \rangle$ per CZ gate under the constraint of 180-degree conditional phase, as described in Appendix~\ref{app:2qb_czgate_calibration}.

\subsection{CZ Gate Benchmarking and Error Analysis}

Following the two-qubit gate calibration, we benchmarked the CZ gate using Clifford-interleaved randomized benchmarking (RB)~\cite{Magesan2012_InterleavedRB}. We obtained an average gate fidelity of {99.64(6)\%} for a \SI{70}{ns} gate time [\fref{fig:cz_rb}(c)]. For this \SI{70}{ns} CZ gate, we performed RB repeatedly over 17 hours, recalibrating the flux biases every half-hour to compensate for flux drift. This long-term measurement yielded an average fidelity of 99.51(6){\%} [\fref{fig:cz_rb}(e)].

We also assessed the CZ gate error as a function of gate time, varying it from \SI{40}{ns} to \SI{100}{ns} [\fref{fig:cz_rb}(d)]. For gate times shorter than \SI{70}{ns}, the gate error increased rapidly, implying a significant rise in coherent errors. Within this gate time, we have shown that fidelity is beginning to be limlited by leakage errors from leakage randomized benchmarking results, which are detailed in \aref{app:2qb_error_analysis}. Conversely, for gate times equal to or longer than \SI{80}{ns}, the gate performance is primarily limited by decoherence. 

Finally, we also measured CZ gate fidelity of device B. Due to its stronger interaction strength, we were able to implement the CZ gate with a shorter duration, ranging from \SI{20}{ns} to \SI{50}{ns}. After finding the idling point, we built the CZ gate using a cosine-shaped flux pulse. We obtained a CZ gate fidelity of 99.68(8)\% at \SI{20}{ns} gate time through Clifford-interleaved RB. The results are summarized in  \fref{fig:tft1_deviceB_summary} of Appendix~\ref{app:tft1_summary}.

\section{\label{sec:discussion_conclusion} CONCLUSION AND DISCUSSION}

In conclusion, we have demonstrated that the TFT architecture can cancel unwanted $ZZ$ interactions while remaining outside of the straddling regime. We also implemented a coupler-flux-biased CZ gate in our TFT system, achieving average gate fidelities over 99.6\% for two different devices with gate times of \SI{70}{ns} and \SI{20}{ns}, respectively. In addition, we achieved gate fidelities exceeding 99\% with gate times ranging from \SI{40}{ns} to \SI{100}{ns}, thereby showing the robustness of the system and the calibration.

The complete cancellation of $ZZ$ interaction is beneficial for achieving robust simultaneous operation of multiple single-qubit gates. As single-qubit gate errors become smaller, even a small amount of residual $ZZ$ interaction can become a dominant error source. Furthermore, in an all-transmon system with a coupler, the $ZZ$ interaction grows with the fourth order of the interaction strength ($\zeta_{ZZ} \propto g_{qc}^{4}$) \cite{FeiYan2021_TTTCZgatetheory}. This quartic dependence can limit the speed of two-qubit gates, as the need to preserve the fidelity of single-qubit gates imposes limitations on the residual $ZZ$ interactions. The TFT system has a significant advantage in this regard because $ZZ$-interaction-free point is achievable regardless of the interaction strength. 

A larger interaction strength is desirable for faster adiabatic gate operation, as it results in a larger avoided crossing that exponentially reduces non-adiabatic leakage for a fixed traversal rate. In device B, which has an interaction strength twice that of device A, we achieved a high-fidelity CZ gate in a duration as short as \SI{20}{ns} without significantly degrading the adiabaticity. 

Finally, we anticipate ways to improve performance with relatively minor modifications to the architecture presented here. For instance, replacing the tunable transmon with a fixed-frequency transmon, which generally has a longer coherence time, would likely result in a considerable reduction in incoherent error. Furthermore, eliminating the tunable transmon could simplify the calibration process by alleviating the need for calibration of flux transients and crosstalk. 

\section{\label{sec:acknowlegements} ACKNOWLEDGEMENTS}
We gratefully acknowledge Anuj Aggarwal for fruitful discussions.
This research was sponsored by the Army Research Office and was accomplished under Award Number: W911NF-23-1-0045 (Extensible and Modular Advanced Qubits) under contract number DE-SC0012704; in part by the U.S. Department of Energy, the Office of Science National Quantum Information Science Research Center’s Co-design Center for Quantum Advantage (Con-
tract No. DE-SC0012704); and in part under Air Force Contract No. FA8702-15-D-0001.
%The U.S. Government is authorized to reproduce and distribute reprints for Government purposes notwithstanding any copyright notation herein.
J.A. and J.K. gratefully acknowledges the support from Korea Foundation for Advanced Studies. M.H. is supported by an appointment to the Intelligence Community Postdoctoral Research Fellowship Program at the Massachusetts Institute of Technology administered by Oak Ridge Institute for Science and Education (ORISE) through an interagency agreement between the U.S. Department of Energy and the Office of
the Director of National Intelligence (ODNI).
The views and conclusions contained herein are those of the authors and should not be interpreted as necessarily representing the official policies or endorsements, either expressed or implied, of the U.S. Government.

% \nocite{*}
\clearpage 
\bibliography{aipsamp2}% Produces the bibliography via BibTeX.
\clearpage
\newpage 
\appendix

\section{Measurement Setup}
\label{appendix:measurement_setup}
The TFT device was measured in a Leiden CF-450 dilution refrigerator, operating at a base temperature of around \SI{20}{mK}. The chip was wire-bonded inside an aluminum package made by MIT Lincoln Laboratory. The chip was magnetically shielded by a superconducting can and a Mu-metal covering.

The readout input and qubit drive signals were generated by an R\&S SGS100A SGMA RF signal generator, acting as a local oscillator, and pulse-modulated by a Keysight M3202 PXIe Arbitrary Waveform Generator (AWG). To suppress thermal noise, the RF signal was attenuated by \SI{66}{dB} to \SI{70}{dB} between room temperature and the device. To further reduce noise, high-pass filters, low-pass filters, or Eccosorb filters were placed after the attenuators at the mixing chamber stage.

The readout output signal was first amplified by Josephson traveling-wave parametric amplifier (JTWPA or TWPA) located at the mixing chamber stage. JTWPA pump signal was first generated by a Holzworth HS9001B RF synthesizer and pulse-modulated by the Keysight M3202 PXIe AWG, ensuring the pump was active only during the readout. After the JTWPA, the readout output signal was further amplified by two high-electron-mobility-transistor (HEMT) amplifiers at the \SI{3}{K} stage and at room temperature. The readout signal was then downconverted by a mixer and amplified by a Stanford Research SR445A preamplifier. Finally, the amplified, downconverted signal was digitized by a Keysight M3102A Analog-to-Digital Converter (ADC or Digitizer).

DC flux signals were generated by Yokogawa GS200 and Yokogawa 7651 DC sources. At room temperature, these DC signals were combined with a baseband flux pulse from a Keysight M3202A AWG. The combined signals were sent to the flux lines through coaxial cables, which had a total attenuation of \SI{26}{dB}, including \SI{0}{dB} attenuators at the mixing chamber stage. Finally, \SI{1}{GHz} low-pass filters (VLFG-1000) and Eccosorb filters were installed after \SI{0}{dB} attenuators to reduce thermal noise and infrared radiation.

\begin{figure*}[t]
    \centering
    \includegraphics[width=\textwidth]{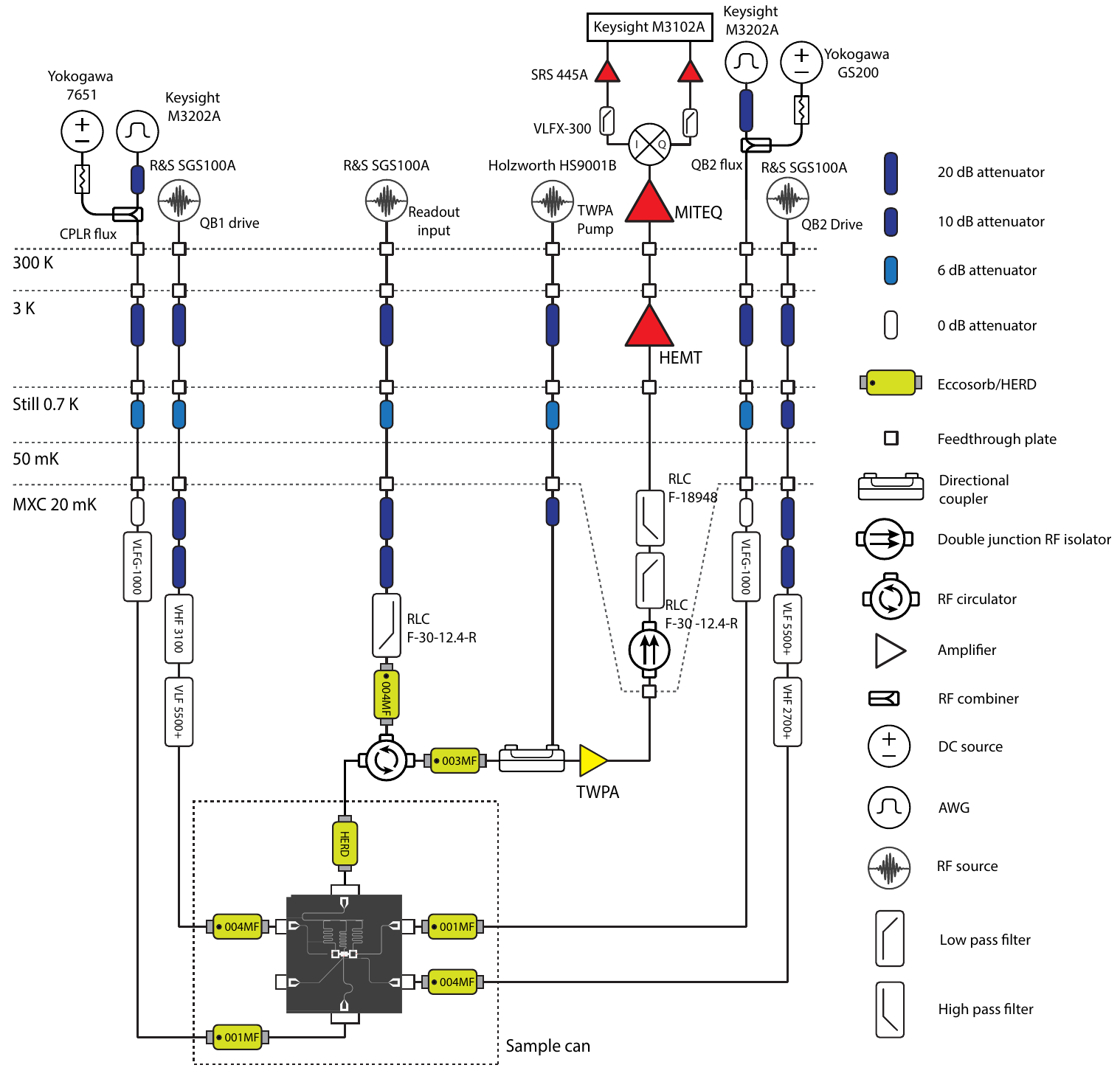}
    \caption{Room-temperature and cryogenic setup used to measure TFT device. In the figure, ``001MF", ``004MF" refer to the cryogenic eccosorb (QMC-CRYOIRF-001MF, 004MF) manufactured by Quantum Microwave.}
    \label{fig:cryogenic_setup}
\end{figure*}

The complete wiring diagram and details of the measurement setup are shown in \fref{fig:cryogenic_setup}. The instruments are summarized in Table~\ref{tab:tft_instruments}.

\begin{table}[H]
    \centering
    \begin{tabular}{c|c|c} 
        \hline 
        \hline
        Instrument type & Manufacturer & Model \\
        \hline 
        DC source (to qubit 2) & Yokogawa & GS200 \\
        DC source (to coupler) & Yokogawa & 7651 \\
        RF source (to chip) & Rhode and Schwarz & SGS100A \\
        RF source (to TWPA) & Holzworth & HS9001B \\ 
        Control chassis & Keysight & M9019A \\
        AWG & Keysight & M3202A \\ 
        ADC & Keysight & M3102A \\
         \hline
         \hline
    \end{tabular}
    \caption{Instruments used in this research.}
    \label{tab:tft_instruments}
\end{table}

\clearpage

\section{TFT Circuit Hamiltonian from Capacitances}
\label{app:tft_circuit_hamiltonian}

\begin{centering}
\begin{figure}[H]
    \centering
    \includegraphics[width=0.45\textwidth]{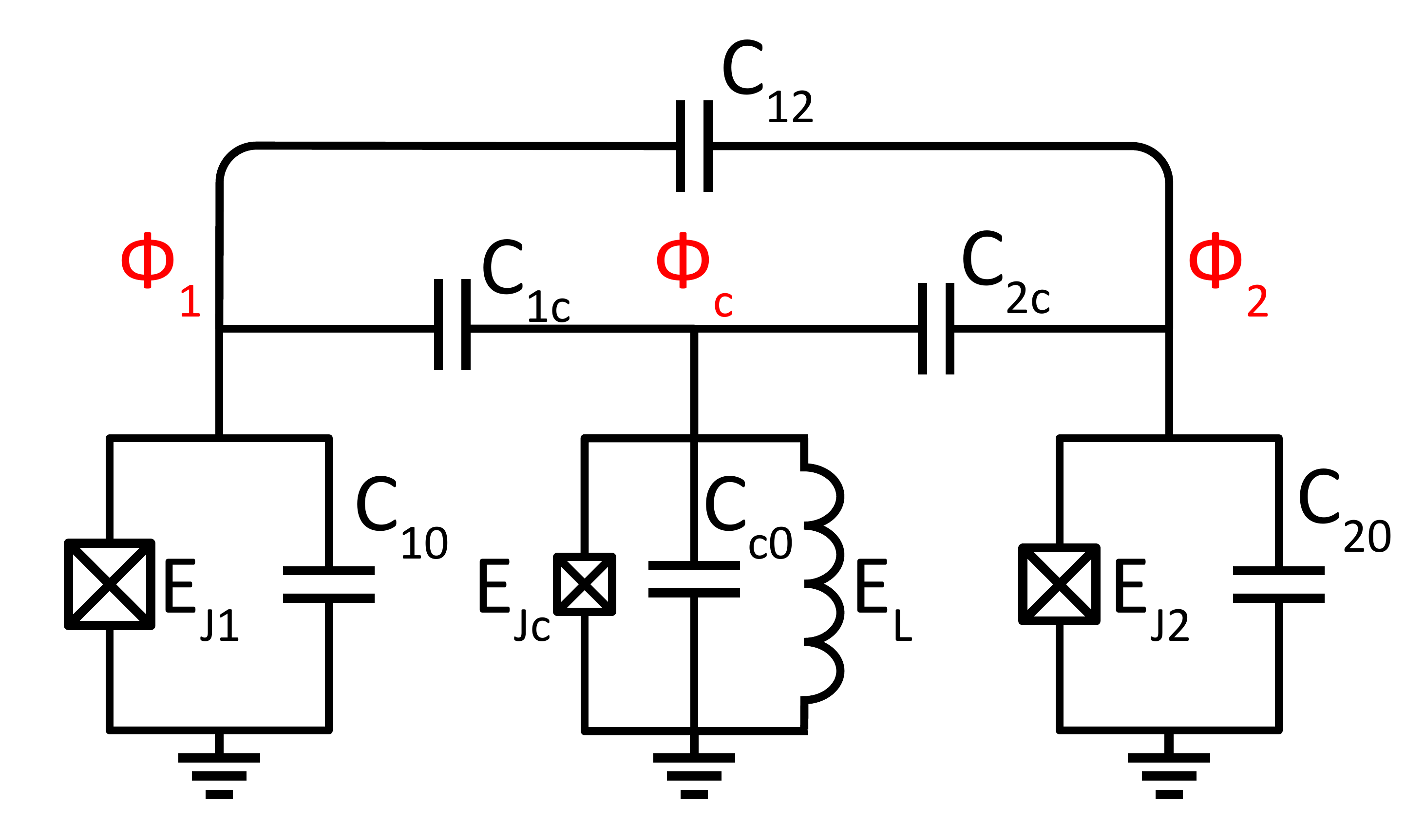}
    \caption{TFT circuit with nodal phase variables $\Phi_{1},\Phi_{c},\Phi_{2}$ and capacitances $C_{ij}$. In the figure, $C_{i0}$ includes the junction capacitance and the shunting capacitance.}
    \label{fig:tft_circuit_appb}
\end{figure}
\end{centering}

In this section, we derive the TFT system Hamiltonian from a circuit diagram using circuit quantization theory. Based on \fref{fig:tft_circuit_appb}, the circuit Lagrangian of the TFT system can be written as \eref{eqn:tft_lagrangian_appb}.

\begin{align} \label{eqn:tft_lagrangian_appb} 
\begin{split} 
    \mathcal{L} & =  \frac{1}{2} C_{10} \dot{\Phi}_{1}^{2} + \frac{1}{2} C_{c0} \dot{\Phi}_{c}^{2} + \frac{1}{2} C_{20} \dot{\Phi}_{2}+ \frac{1}{2} C_{1c} (\dot{\Phi}_{1} - \dot{\Phi}_{c})^{2} \\ 
    &+ \frac{1}{2} C_{2c} (\dot{\Phi}_{c} - \dot{\Phi}_{2})^{2} + \frac{1}{2} C_{12} (\dot{\Phi}_{1} - \dot{\Phi}_{2})^{2} - \frac{1}{2} E_{L} \varphi_{c}^{2} \\ 
    & + E_{J1} \cos \varphi_{1}   + E_{J2} \cos \varphi_{2} + E_{Jc} \cos (\varphi_{c} - \varphi_{\textrm{ext}})
\end{split}
\end{align} where $\varphi_{k} = 2\pi \Phi_{k}/\Phi_{0}$. The Lagrangian above can be simplified as below.
\begin{align}
\begin{split}
    \mathcal{L} & = \frac{1}{2} \dot{\mathbf{\Phi}}^{T} [\mathbf{C}] \dot{\mathbf{\Phi}} - \frac{1}{2} E_{L} \varphi_{c}^{2} \\
    &+ E_{J1} \cos \varphi_{1} + E_{J2} \cos \varphi_{2} + E_{Jc} \cos (\varphi_{c} - \varphi_{\textrm{ext}})
\end{split}
\end{align}
where $\mathbf{\Phi}^{T} = [\Phi_{1},\Phi_{c},\Phi_{2}]$ and the capacitance matrix $[\mathbf{C}]$ has the following expression.
\begingroup\makeatletter\def\f@size{9}\check@mathfonts
\def\maketag@@@#1{\hbox{\m@th\large\normalfont#1}}%
\begin{align}
    [\mathbf{C}] = \begin{bmatrix} C_{10} + C_{1c} + C_{12} & -C_{1c} & -C_{12} \\ 
   -C_{1c} & C_{c0} + C_{1c} + C_{2c} & -C_{2c} \\ 
   -C_{12} & -C_{2c} & C_{20} + C_{12} + C_{2c} \end{bmatrix}
\end{align} \endgroup

To derive the Hamiltonian of the system, we perform Legendre transformation $H = \sum_{k=\{1,c,2 \}} \dot{\Phi}_{k} Q_{k} - \mathcal{L} $ where $Q_{k} \equiv \partial \mathcal{L} / \partial \dot{\Phi}_{k}$. By defining $[Q_{1},Q_{c},Q_{2}]^{T} \equiv \mathbf{Q}  = [\mathbf{C}] \dot{\mathbf{\Phi}}$, we get the following result:
\begin{align}
\begin{split}
    H &= \frac{1}{2} \mathbf{Q}^{T} [\mathbf{C}]^{-1} \mathbf{Q} + \frac{1}{2} E_{L} \varphi_{c}^{2} \\
    & - E_{J1} \cos \varphi_{1} - E_{J2} \cos \varphi_{2} - E_{Jc} \cos (\varphi_{c} - \varphi_{\textrm{ext}}). 
\end{split}
\end{align}

\noindent $Q_{k}$ has a dimension of charge, thus we can define cooper pair number operator $n_{k} = Q_{k}/2e$ and write the Hamiltonian in a more familiar form,
\begin{align}
\begin{split}
    H &=  4E_{C1} n_{1}^{2} + 4E_{Cc} n_{c}^{2} + 4E_{C2} n_{2}^{2}  + \frac{1}{2} E_{L} \varphi_{c}^{2} \\
    & + J_{1c} n_{1}n_{c} + J_{2c} n_{c}n_{2} + J_{12}n_{1}n_{2} \\
    & - E_{J1} \cos \varphi_{1} - E_{J2} \cos \varphi_{2} - E_{Jc} \cos (\varphi_{c} - \varphi_{\textrm{ext}}),
\end{split}
\end{align}
where each energy parameter is defined as
\begin{align}
\begin{split}
    E_{C1} &= {e^{2}} [\mathbf{C}]^{-1}[0,0] /2 \\
    E_{Cc} &= {e^{2}} [\mathbf{C}]^{-1}[1,1] /2 \\
    E_{C2} &= {e^{2}} [\mathbf{C}]^{-1}[2,2] /2 \\
    J_{1c} &= 4 e^{2} [\mathbf{C}]^{-1}[0, 1] \\
    J_{2c} &= 4e^{2} [\mathbf{C}]^{-1}[1,2] \\
    J_{12} &= 4e^{2} [\mathbf{C}]^{-1}[0,2]. 
\end{split}
\end{align}

The experimentally measured energy parameters are listed in Table~\ref{tab:tft_param} of the main text.

\section{Qubit Readout}
\label{app:qubit_readout_parameters}
\begin{center}

\begin{figure}[H]
    \centering
    \includegraphics[width=0.98 \linewidth]{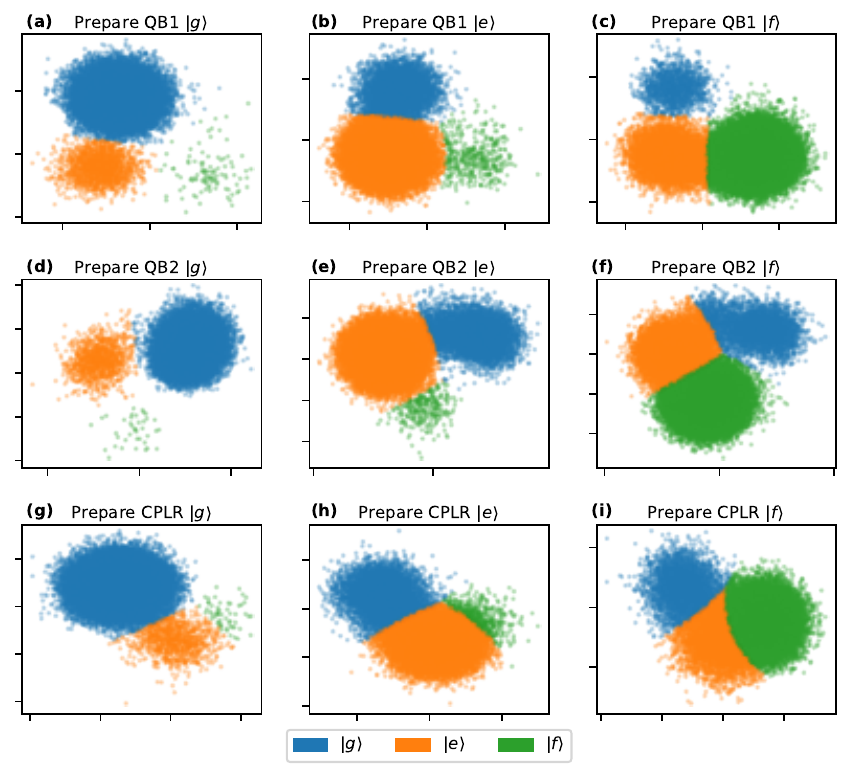}
    \caption{State assignment scatter plot for the first three states of qubit 1, qubit 2, and coupler in device A. The states were discriminated by fitting the single-shot results using a Gaussian mixture model. The first-excited states were prepared by a $\pi$-pulses calibrated through Rabi oscillations, and the second-excited states were prepared by two-photon transitions.}
    \label{fig:state_assignment}
\end{figure}
\end{center}

The readout parameters for devices A and B are listed in Table~\ref{tab:tft_ro_parameters}. Additionally, the state assignment fidelities for the first three states of each qubit are summarized in Table~\ref{tab:state_assignment_fidelity} and illustrated in \fref{fig:state_assignment}.

\begin{center}
\begin{table}[H]
    \begin{tabular}{>{\centering}m{1.0cm} | >{}p{5.4cm}|  >{\centering\arraybackslash}m{1.5cm}}
        \hline \hline
      Device & Parameters & Value \\
         \hline
      A & Qubit 1 bare resonator frequency &  \SI{6.973}{GHz} \\
       & Qubit 2 bare resonator frequency&  \SI{6.833}{GHz} \\
       & Coupler bare resonator frequency&  \SI{7.111 }{GHz} \\
       & Qubit 1 0-1 dispersive shift ($\chi_{01}/2\pi$) &  \SI{0.43}{GHz} \\
       & Qubit 2 0-1 dispersive shift ($\chi_{01}/2\pi$) &  \SI{0.57}{GHz} \\
      & Qubit 1 resonator linewidth ($\kappa/2\pi$) &  \SI{0.63}{MHz} \\
      & Qubit 2 resonator linewidth ($\kappa/2\pi$) &  \SI{0.69}{MHz} \\
      \hline
      B & Qubit 1 bare resonator frequency &  \SI{7.081}{GHz} \\
       & Qubit 2 bare resonator frequency&  \SI{6.972}{GHz} \\
       & Coupler bare resonator frequency&  \SI{7.233 }{GHz} \\
       & Qubit 1 0-1 dispersive shift ($\chi_{01}/2\pi$) &  \SI{0.37}{GHz} \\
       & Qubit 2 0-1 dispersive shift ($\chi_{01}/2\pi$) &  \SI{0.33}{GHz} \\
      & Qubit 1 resonator linewidth ($\kappa/2\pi$) &  \SI{0.511}{MHz} \\
      & Qubit 2 resonator linewidth ($\kappa/2\pi$) &  \SI{0.527}{MHz} \\
      
       \hline \hline
    \end{tabular}
    \caption{Bare readout resonator frequencies ($\omega_{r}^{0}/2\pi$), resonant frequency shift between ground and first-excited states ($\chi_{01}/2\pi$), and linewidth ($\kappa / 2 \pi$) for qubit 1, qubit 2, and the coupler at zero $ZZ$ interaction flux bias.}
    \label{tab:tft_ro_parameters}
    \end{table}
\end{center}

\begin{center}
\begin{table}[H]
    \centering
    \begin{tabular}{>{\centering}m{2.5cm} >{\centering}m{0.5cm}| >{\centering}m{1.5cm} >{\centering}m{1.5cm} >{\centering\arraybackslash}m{1.5cm} }
    \hline 
    \hline
   Qubit 1 & & \multicolumn{3}{c}{Prepared states}  \\
    && $|0 \rangle $ & $|1 \rangle$ & $|2 \rangle $ \\
    \hline
   & $|0\rangle $ & 0.967 & 0.066 & 0.016 \\
  Assigned states & $|1 \rangle$ & 0.032 & 0.930 & 0.101 \\
   & $|2 \rangle$ & 0.001 & 0.004 &  0.884 \\
   \hline 
   \hline
   Qubit 2 & & \multicolumn{3}{c}{Prepared states}  \\
    && $|0 \rangle $ & $|1 \rangle$ & $|2 \rangle $ \\
    \hline
   & $|0\rangle $ & 0.974 & 0.084 & 0.044 \\
  Assigned states & $|1 \rangle$ & 0.025 & 0.913 & 0.204\\
   & $|2 \rangle$ & 0.001 & 0.003 & 0.752 \\
   \hline 
   \hline
   Coupler & & \multicolumn{3}{c}{Prepared states}  \\
    && $|0 \rangle $ & $|1 \rangle$ & $|2 \rangle $ \\
    \hline
   & $|0\rangle $ & 0.987 & 0.156 & 0.076 \\
  Assigned states & $|1 \rangle$ & 0.012 & 0.788 & 0.141\\
   & $|2 \rangle$ & 0.000 & 0.057 & 0.783 \\
   \hline 
   \hline   
    \end{tabular}
    \caption{State assignment fidelities of the first three states of qubit 1, qubit 2, and coupler in device A. The assignment fidelities are obtained by measuring $P(m|n)$, the probability we assign $|m \rangle$ when we prepare state $|n \rangle$. }
    \label{tab:state_assignment_fidelity}
\end{table}
\end{center}

\section{Coherence of a TFT System}
\label{app:qubit_coherence_repeat}

\begin{figure*}[t]
    \centering
    \includegraphics[width=0.95\linewidth]{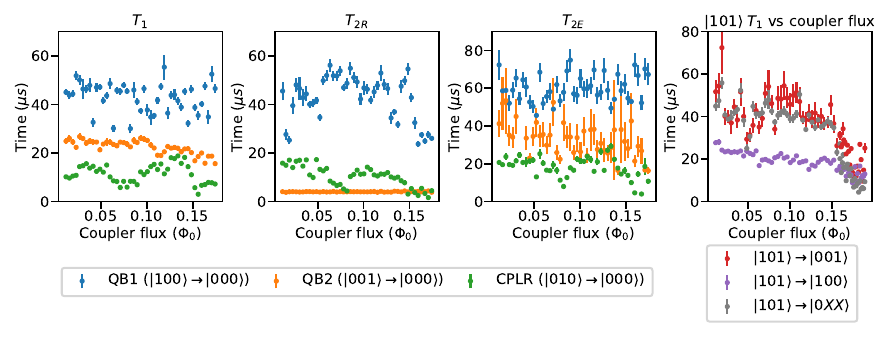}
    \caption{$T_{1}$, $T_{2R}$, $T_{2E}$ of qubit 1, qubit 2, and the coupler as a function of coupler flux (device A). The coupler flux covers the range of CZ gate operation, from the zero-$ZZ$ bias to the avoided crossing between $|101 \rangle$ and $|030 \rangle$. In this figure, $|0 XX \rangle$ includes all the states that can transition from $|101 \rangle$ except $|100\rangle$. $T_{1}$ between $|101 \rangle $ and $|001\rangle, |100\rangle$ are calculated by post-selecting the single-shot results of the decay from $|101 \rangle$. $T_{1}$ between $|101 \rangle$ and $|0XX \rangle$ is calculated by fitting the readout signal measured in qubit 1 resonator without any post-selection. The difference between $T_{1}^{|101 \rangle \rightarrow |001\rangle}$ and $T_{1}^{|101 \rangle \rightarrow |0XX \rangle}$ can be considered the relaxation time from $|101 \rangle$ to $|000 \rangle, |010 \rangle, |020 \rangle, |030 \rangle$.}

    \label{fig:TFT_coherence_vs_flux}
\end{figure*}

In this section, we present the coherence times for device A. First, we measured the relaxation ($T_{1}$), Ramsey dephasing ($T_{2}^{R}$), and Hahn-echo dephasing ($T_{2}^{E}$) times of the first two states of qubit 1, qubit 2, and the coupler over the operational flux range of the CZ gate. In addition, we measured $T_{1}$ of $|101\rangle$ in the same flux range to estimate the incoherent error limit of the CZ gate more accurately. Hybridization between $|101 \rangle$ and $|030 \rangle$ near their avoided crossing allows direct relaxation from $|101 \rangle$ to $|000 \rangle$, which in turn lowers the average $T_{1}$ used to calculate the the incoherent error limit [ \eref{eqn:2qberr_incoh_limit}]. The results are summarized in \fref{fig:TFT_coherence_vs_flux}.

\begin{figure}[H]
    \centering
    \includegraphics[width=0.99 \linewidth]{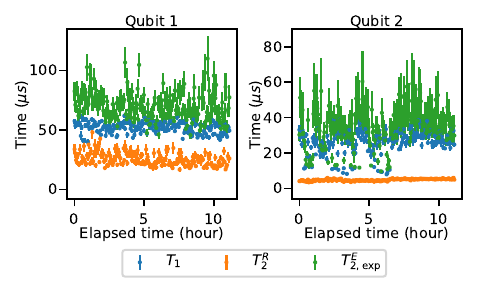}
    \caption{Repeated coherence time measurement of qubit 1 and qubit 2 at zero-$ZZ$ flux bias. The coherence times are measured for 12 hours.}
    \label{fig:TFT_coherence_repeat}
\end{figure}

Next, we repeatedly measured the $T_{1}$, $T_{2}^{R}$, $T_{2}^{E}$ times of the first two states of qubit 1 and qubit 2 at the idle point over a 12-hour period [\fref{fig:TFT_coherence_repeat}]. Fluctuations in these coherence times can be used to estimate the lower and upper bounds of the incoherent error. The measurement results are sumamrized in Table~\ref{tab:tft_coherence_time_app}.

\section{Flux-Noise Amplitude Measurement}
We calculated the flux-noise amplitudes of the coupler and qubit 2 by measuring the Hahn-echo dephasing time at qubit frequencies with different flux sensitivities ($d f / d\Phi_{\textrm{ext}}$) following the procedure introduced in \cite{Braumuller2020_FluxNoiseAmp}. We fit the data from the Hahn-echo sequence using an exponential Gaussian decay $A e^{-t/T_{2,\textrm{exp}}^{E}}e^{(-t/T_{\varphi }^{E})^{2}} + B$, and calculated the flux-noise amplitude $\sqrt{A_{\Phi}}$ using equation $1/T_{\varphi}^{E} = 2\pi \sqrt{A_{\Phi} \ln2} |d f / d \Phi_{\textrm{ext}}|$.

\begin{figure}[H]
    \centering
    \includegraphics[width=0.99\linewidth]{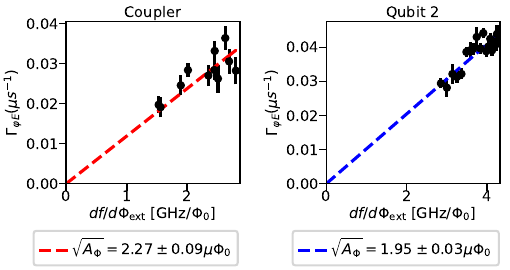}
    \caption{Flux-noise amplitude measurement results of the coupler and qubit 2 in device A.}
    \label{fig:tft_flux_noise_amp}
\end{figure}

The resulting flux-noise amplitudes of the coupler and qubit 2 in device A were 2.27$\pm$$\SI{0.09}{\micro \Phi_{0}}$ and 1.95$\pm$$\SI{0.03}{\micro \Phi_{0}}$, respectively [\fref{fig:tft_flux_noise_amp}].

\section{Single-Qubit Gate Error Analysis}
\label{app:single_qubit_coherence_and_error}

We estimate the incoherent error of a single-qubit gate $r_{\textrm{1Q}}$ using \eref{eqn:incoherent_error_1qb}, which includes the decoherence caused by the $T_{1}$ decay, white noise dephasing, and $1/f$ flux-noise dephasing \cite{O'Malley2015_SingleQubitRB}:

\begin{align} \label{eqn:incoherent_error_1qb}
    r_{\textrm{1Q}}(t) = \frac{t}{3T_{1}} + \frac{1}{6} \left(  \langle \Delta \phi_{1/f}^{2} (t) \rangle + \langle \Delta \phi_{\textrm{white}}^{2}(t) \rangle \right).
\end{align}

In \eref{eqn:incoherent_error_1qb}, $ \langle \Delta \phi^{2} \rangle $ is the variance of the random phase error caused by the noise source. For instance, white-noise-induced phase variance $\langle \Delta \phi_{\textrm{white}}^{2} \rangle$ satisfies 
\begin{align} \label{eqn:white_noise_phase_variance}
    \frac{1}{6}\langle \Delta \phi_{\textrm{white}}^{2}(t) \rangle = \frac{t}{3 T_{\varphi,\textrm{exp}}^{E}} = \frac{t}{3 T_{2,\textrm{exp}}^{E}} - \frac{t}{2 T_{1}}.
\end{align}
Similarly, $1/f$-flux-noise-induced phase variance $\langle \Delta \phi_{1/f}^{2} (t) \rangle$ satisfies
\begin{align} \label{eqn:inverse_fnoise_phase_variance}
\begin{split}
\langle \Delta \phi_{1/f}^{2} (t) \rangle & = 4 A_{\Phi} \left(\frac{\partial \omega}{\partial \Phi_{\textrm{ext}}} \right)^{2} \int_{\omega_{\textrm{IR}}}^{\omega_{\textrm{UV}}} \frac{1 - \cos \omega t } {\omega^{3}}  d \omega \\ 
& = \frac{4}{ (T_{\varphi}^{E})^{2} \ln 2} \int_{\omega_{\textrm{IR}}}^{\omega_{\textrm{UV}}} \frac{1 - \cos \omega t } {\omega^{3}}  d \omega,
\end{split}
\end{align}
where $D_{1} = \partial \omega / \partial \Phi_{\textrm{ext}}$, and $\omega_{\textrm{IR}}/2\pi$, $\omega_{\textrm{UV}}/2\pi$ are low-cutoff and high-cutoff frequencies. In this measurement, we set the ultraviolet (UV) cutoff $\omega_{\textrm{UV}}/2\pi = $\SI{100}MHz, corresponding to the frequensy scale of a \SI{10}ns gate time, and the infrared (IR) cutoff to $\omega_{\textrm{IR}}/2\pi=$\SI{1}Hz, corresponding to the inverse of the time required to measure a single datapoint. Substituting \eref{eqn:white_noise_phase_variance}, \eref{eqn:inverse_fnoise_phase_variance} to \eref{eqn:incoherent_error_1qb}, we have the following expression:
\begin{align} \label{eqn:1QB_error_total}
    r_{\textrm{1Q}} = \frac{t}{3} \left(\frac{1}{2T_{1}} + \frac{1}{T_{2,\textrm{exp}}^{E}} \right) + \frac{2}{3(T_{\varphi}^{E})^{2} \ln 2} \int_{\omega_{\textrm{IR}}}^{\omega_{\textrm{UV}}} \frac{1 - \cos \omega t } {\omega^{3}}  d \omega. 
\end{align}

Table~\ref{tab:tft_coherence_time_app} summarizes the theoretical and measured incoherent errors, along with the coherence times at the zero-$ZZ$ flux bias.

\begin{center}
\begin{table*}[t]
    \begin{tabular}{>{\centering}p{1.0cm} >{\centering}m{1.0cm} >{\centering}m{1.6cm} >{\centering}m{1.6cm} >{\centering}m{1.6cm} >{\centering}m{1.6cm}>{\centering}m{1.6cm} >{\centering}m{2.6cm}>{\centering\arraybackslash}m{2.6cm} }
        \hline \hline
        Device & Qubit & $T_{1}$(\SI{}{\micro s}) &  $T_{2}^{R}$ (\SI{}{\micro s})& $T_{2,\textrm{exp}}^{E}$(\SI{}{\micro s})  & $T_{\varphi}^{E}$ (\SI{}{\micro s})& 
        $\langle \Delta \phi_{1/f}^{2} \rangle$ &
        $r_{\textrm{1Q}}^{\textrm{theory}}$ & $r_{\textrm{1Q}}^{\textrm{meas}}$ \\
         \hline
        A & QB1  & 51.3$\pm$4.4 & 25.2$\pm$5.3 & 70.0$\pm$11.6 & - & - &(4.0$\pm$0.5)$\times 10^{-4}$ & (6.6$\pm$0.2)$\times 10^{-4}$ \\
         % & QB1 (QB2=$|1\rangle$) & 25.0 & 6 & 11 & 14.1  & -  &$1.8 \times 10^{-3}$ &  \\
         & QB2  & 25.0$\pm$6.9 & 4.8$\pm$0.6 & 34.6$\pm$11.0  & 22.3$\pm$3.3 & $2.3 \times 10^{-4}$ &(8.5$\pm$2.4)$\times 10^{-4}$ & (1.0$\pm$0.0)$\times 10^{-3}$  \\
         % & QB2 (QB1=$|1\rangle$) & & & & & & & \\
         \hline
         B & QB1 & 23.0$\pm$8.5 & 27.4$\pm$7.5 & 31.0$\pm$10.7 & - & - & (1.1$\pm$0.4)$\times 10^{-3}$ & (1.7$\pm$0.4)$\times 10^{-3}$ \\
         & QB2 & 45.8$\pm$13.7 & 5.4$\pm$1.1 & 75.1$\pm$11.6 & 14.7$\pm$0.6 & $7.1 \times 10^{-4}$ & (6.1$\pm$2.3)$\times 10^{-4}$ & (1.1$\pm$0.2)$\times 10^{-3}$ \\ 
         \hline \hline
    \end{tabular}
    \caption{Average coherence times and corresponding incoherent error limit of single-qubit gates for device A and device B. Coherence times were measured at the zero-$ZZ$ flux bias. The uncertainty of the coherence time represents the standard deviation of the multiple measurement.}
    \label{tab:tft_coherence_time_app}
    \end{table*}
\end{center}

\section{Two-Qubit Gate Error Analysis}
\label{app:2qb_error_analysis}

In a two-qubit gate, we use the following equation to estimate the incoherent error limit \cite{LeonDing2023_FTF, RuiLi2024DTSCZGate, Pedersen2007_incoherentlimit}:

\begin{align} \label{eqn:2qberr_incoh_limit}
\begin{split}
    r_{CZ}^{\textrm{incoh}} &= \frac{2t_{CZ}}{5} \Big( \frac{1}{\bar{T}_{1}^{\mathrm{QB1}}}  + \frac{1}{\bar{T}_{1}^{\mathrm{QB2}}}  +  \frac{1}{\bar{T}_{\varphi,\textrm{exp}}^{E,\textrm{QB1}}}+ \frac{1}{\bar{T}_{\varphi,\textrm{exp}}^{E, \textrm{QB2}}} \Big), \\
\end{split}
\end{align}
where $\bar{T}$ denotes the average coherence time over the CZ gate duration. However, using the coherence time measured at the zero-$ZZ$ flux bias would underestimate the incoherent error, as the coherence time decreases when the system moves away from this point.

Therefore, to establish a more realistic lower bound of the incoherent error, we use the $T_{1}$ as a function of flux from \fref{fig:TFT_coherence_vs_flux} to calculate the averaged $T_{1}$ during the CZ gate. Additionally, to account for state-dependent decay, we use the following modified equation to estimate the $T_{1}$-like error during the CZ gate \cite{FeiYan2021_TTTCZgatetheory}:

\fontsize{7}{7}\selectfont
\begin{align}
\begin{split}
    &r_{CZ}^{T_{1}} = \frac{t_{CZ}}{5} \Big( \frac{1}{\bar{T}_{1}^{\mathrm{100 \rightarrow 000}}}  + \frac{1}{\bar{T}_{1}^{001 \rightarrow 000}}  + \frac{1}{\bar{T}_{1}^{101 \rightarrow001}} + \frac{1}{\bar{T}_{1}^{101 \rightarrow 100}} + \frac{1}{\bar{T}_{1}^{101 \rightarrow 000}}  \\ 
    & + \frac{1}{\bar{T}_{1}^{100 \rightarrow 001}} + \frac{1}{\bar{T}_{1}^{001 \rightarrow 100}}\Big )+ \frac{t_{CZ}}{4} \left( \frac{1}{ \bar{T}_{1}^{101 \rightarrow }} + \frac{1}{\bar{T}_{1}^{100 \rightarrow }} + \frac{1}{\bar{T}_{1}^{001 \rightarrow }} \right),  
\end{split}
\end{align}
\normalsize
where $1/\bar{T}_{1}^{101 \rightarrow}$ denotes the transition rate from $|101 \rangle$ state to non-computational states, and similar for $1/\bar{T}_{1}^{100 \rightarrow}$, $1/\bar{T}_{1}^{001 \rightarrow}$. 

From the above equation, we can write a lower bound of the incoherent CZ gate error as 

\fontsize{8}{8}\selectfont
\begin{align}
\begin{split}
    r_{CZ}^{\textrm{incoh}} & \geq  \frac{t_{CZ}}{5} \Big[ \frac{1}{\bar{T}_{1}^{\mathrm{100 \rightarrow 000}}}  + \frac{1}{\bar{T}_{1}^{001 \rightarrow 000}}  + \frac{1}{\bar{T}_{1}^{101 \rightarrow 100}}  \\ 
    & +\frac{1}{T_{1}^{101 \rightarrow 0XX }} + \frac{2}{\bar{T}_{\varphi,\textrm{exp}}^{E,\textrm{100}}}+ \frac{2}{\bar{T}_{\varphi,\textrm{exp}}^{E, \textrm{001}}}  \Big],
\end{split}
\end{align}
\normalsize
where $|0XX \rangle$ includes the state with zero first excitation number: \{$|000\rangle, |010 \rangle, |001 \rangle, |020 \rangle,|030 \rangle$\}, which we measured in \fref{fig:TFT_coherence_vs_flux}. Using the coherence time profile and the CZ gate pulse shape, we estimated the incoherent error bound for each CZ gate time.

Next, the two-qubit gate leakage error was analyzed using the leakage randomized benchmarking (LRB) technique \cite{Wood2018_LRB, RuiLi2024DTSCZGate, RuiLi2025_DTSPRA},  following the steps outlined below.

\begin{itemize}
  \item Prepare a sequence of $m$ random two-qubit Cliffords and the recovery gate at the final (reference sequence). After the reference sequence, count the population remaining in computational subspace, $P_{\textrm{comp}}^{\textrm{ref}}(m)$.
  \item Fit $P_{\textrm{comp}}^{\textrm{ref}}(m)$ to the model $A_{\textrm{ref}} p_{\textrm{ref}}^{m} + B_{\textrm{ref}}$. Calculat the leakage error $L_{1}^{\textrm{ref}} = (1 - p_{\textrm{ref}})(1 - B_{\textrm{ref}})$.
  \item Repeat steps 1 and 2 for an interleaved sequence, where a CZ gate is inserted between each Clifford gate of the reference sequence. The corresponding leakage error $L_{1}^{\mathrm{int}}$ can be determined.  
  \item Calculate the leakage error from a CZ gate, which satisfies  $L_{1}^{\textrm{CZ}} = 1 - \frac{1 - L_{1}^{\textrm{int}}}{1 - L_{1}^{\textrm{ref}}}$. 
  \item Next, measure the ground state population after the reference sequence of $m$ two-qubit Cliffords, $P_{000}^{\textrm{ref}}(m)$ . Fit the quantity $P_{000}^{\textrm{ref}}(m) - P_{\textrm{comp}}^{\textrm{ref}}(m)/4$ to the model $C_{\textrm{ref}}q_{\textrm{ref}}^{m} + D_{\textrm{ref}}$. Repeat for the interleaved sequence to obtain $C_{\textrm{int}}$, $q_{\textrm{int}}$, and $D_{\textrm{int}}$. 
  \item Calculate $q_{CZ} = q_{\mathrm{int}}/q_{\mathrm{ref}}$. The average CZ gate error can be expressed as $r_{CZ}=\frac{3}{4} \left(1  - \frac{q_{\mathrm{int}}}{q_{\mathrm{ref}}} \right)$. 
  \item Finally, calculate the total CZ gate fidelity usiung the expression $F_{\textrm{CZ}}=1 - r_{\textrm{CZ}} - L_{1}^{\textrm{CZ}}/4$. 
\end{itemize}

\begin{figure}[H]
    \centering
    \includegraphics[width=0.98\linewidth]{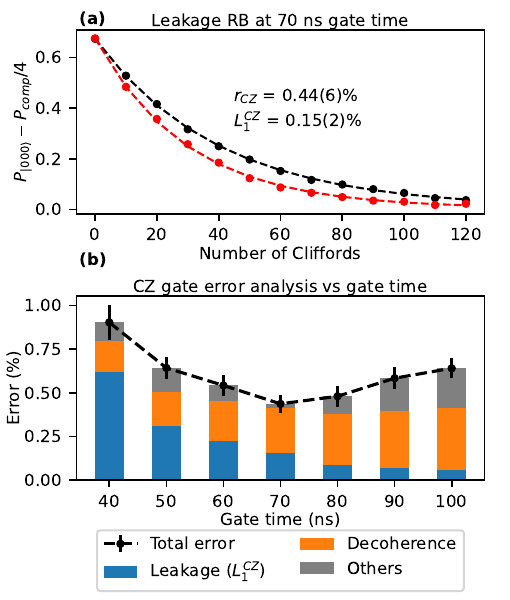}
    \caption{(a) Leakage RB result example at 70 ns gate time. (b) CZ gate error analysis as a function gate time. We see the clear trend of increasing leakage as the gate time decreases.}
    \label{fig:LRB_vs_CZtime}
\end{figure}

The estimated leakage and incoherent error for gate times from \SI{40}{ns} to \SI{100}{ns} are summarized in \fref{fig:LRB_vs_CZtime}. For the range of gate times measured, the leakage RB result suggests that leakage error and decoherence can explain most of the CZ gate error. The gray region of the bar ``others" could consist of fluctuations of coherence time or additional coherent errors from miscalibration, flux fluctuations, and imperfect flux transient calibration.

\section{Flux Transient Calibration}
\label{app:flux_transient_calibration}

In a TFT circuit, a two-qubit CZ gate is implemented by applying flux pulses to the qubits. The flux pulse generated by the AWG undergoes considerable distortion before reaching the qubits. As a result, the qubit would experience a transient flux that persists even after the pulses are nominally turned off. This flux transient can significantly degrade the fidelity of the CZ gate; therefore, the room-temperature flux pulse must be predistorted to compensate for this effect.

The flux transient of a system is modeled as a series of linear time-invariant transfer functions $h_{k}(t)$s. That is, the relation between the input signal, $V_{\textrm{in}}(t)$, and the signal that reaches the qubit, $V_{\textrm{out}}(t)$, can be written as follows:

\begin{align}
    V_{\textrm{out}}(t) = V_{\textrm{in}}(t) \ast h_{1} (t) \ast h_{2} (t) \ast \cdots \ast h_{N}(t). 
\end{align}

\noindent Next, the step-response of each transfer function $s_{k}(t) = h_{k}(t) \ast u(t) $ is analytically modeled as a sum of exponential decay and damped oscillation, following Ref~\cite{Guo2024_PDSTSinusoidal}:

\begin{align}
    s_{k}(t) = u(t) \left[1 + A_{1k}e^{-t/\tau_{1k}} + A_{2k} e^{-t/\tau_{2k}} \cos \left( \frac{2 \pi t}{T} + \phi \right) \right ].
\end{align}

\begin{figure}[H]
    \centering
    \includegraphics[width=0.48\textwidth]{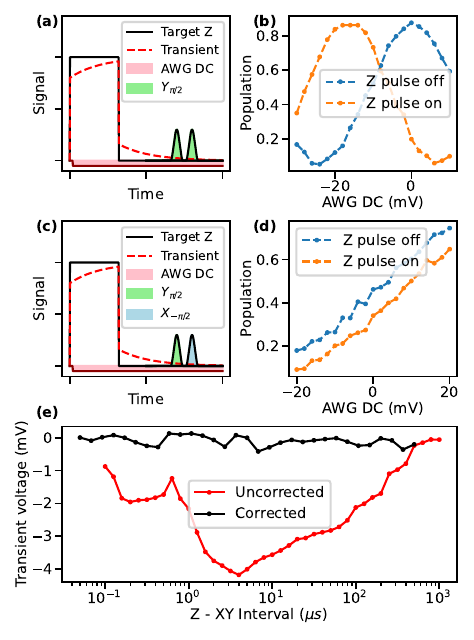}
    \caption{(a) Pulse configuration of $Y_{\pi/2}-Y_{\pi/2}$ after Z and (b) its calibration example. (c) Pulse configuration of $Y_{\pi/2}-X_{-\pi/2}$ after Z and (d) its calibration example. (e) Flux transient from a square pulse before and after the predistortion.} 
    \label{fig:raz_and_rwz}
\end{figure}

To measure the flux transient, we applied a square flux pulse of duration $T$. After a variable delay time $\tau$, we applied single-qubit tomographic pulses $Y_{\pi/2}$, $Y_{\pi/2}$ or $Y_{\pi/2}, X_{-\pi/2}$. Then, we measured the excited-state population as we sweep the DC offset from the AWG. We then compared the population when the flux pulse was turned on and off. The DC offset shift between the two signals are the amount of transient at time $\tau$, which we denote as $V_{\textrm{out}}(\tau)$. Finally, we fit $V_{\textrm{out}}(\tau)$ to a function $s_{k}(\tau + T) - s_{k}(\tau)$ to extract the filter parameters $A_{1k},\tau_{1k},A_{2k},\tau_{2k},T,\phi$. The pulse sequence and example results are shown in \fref{fig:raz_and_rwz}. 

We performed four sets of flux transient calibration before implementing a CZ gate: coupler, qubit 2, coupler to qubit 2, and qubit 2 to coupler. The results are summarized in Table~\ref{tab:flux_transient_calibration}. Using the transient parameters we obtained, the signals arriving at the coupler and qubit 2, denoted $V_{c}^{\textrm{cryo}}(z)$ and $V_{2}^{\textrm{cryo}}(z)$, can be written in the Z domain as below: 

\begin{align}
\begin{split}
    V_{{c}}^{\textrm{cryo}}(z) &= H_{cc}(z) V_{c}^{\mathrm{RT}}(z) + H_{c2}(z) H_{22}(z) V_{2}^{\textrm{RT}}(z) \\ 
    V_{2}^{\textrm{cryo}}(z) &= H_{2c}(z) H_{cc}(z)  V_{c}^{\mathrm{RT}}(z) + H_{22}(z) V_{2}^{\textrm{RT}}(z).
\end{split}
\end{align}

\noindent We can invert this as follows:

\begin{align} \label{eqn:flux_predistortion_2x2}
    \begin{bmatrix}
        V_{c}^{\textrm{RT}}\\ V_{2}^{\textrm{RT}}
    \end{bmatrix}
    = \frac{H_{cc}^{-1}H_{22}^{-1}}{1 - H_{2c} H_{c2}} \begin{bmatrix}
    H_{22} & -H_{c2} H_{22} \\ 
    -H_{2c} H_{cc}& H_{cc}
    \end{bmatrix} 
    \begin{bmatrix}
    V_{c}^{\textrm{cryo}} \\ 
    V_{2}^{\textrm{cryo}}
    \end{bmatrix}.
\end{align}

\noindent Thus, given the desired signals $V_{c}^{\textrm{cryo}}$, $V_{2}^{\textrm{cryo}}$, the corresponding room-temperature signals $V_{c}^{\textrm{RT}}$, $V_{2}^{\textrm{RT}}$ are determined using Eq.~\ref{eqn:flux_predistortion_2x2}. 

\begin{table}[H]
    \centering
    \begin{tabular}{>{\centering}m{1.3cm} >{\centering}m{1.1cm} >{\centering}m{1.1cm} >
    {\centering}m{2.1cm} >
    {\centering\arraybackslash}m{2.1cm}}
        \hline
        \hline
    Device A & CPLR & QB2 & CPLR $\rightarrow$ QB2 & QB2 $\rightarrow$ CPLR\\
        \hline
      $A_{11} $(mV) & 5.1 & 40.0 & -244.6 & 10.7 \\
       $\tau_{11}$(\SI{}{\micro s}) & 160.5 & 126.6 & 184.9 & 283.2\\
       $A_{21} $(mV) & -93.0 & -8.3 & - & - \\
       $\tau_{21}$(\SI{}{\micro s}) & 1.9 & 0.210 & - & - \\
       $T_{1}$(\SI{}{\micro s}) & 489.6 & 169.7 & - & -\\
       $\phi_{1}$ & 1.549 & 0.285 & - & - \\
       \hline
       $A_{12}$(mV) & -27.0 & -61.2 & -19.8 & 1.1 \\
       $\tau_{12}$(\SI{}{\micro s}) & 0.028 & 0.012 & 18.7 & 40.7\\
       $A_{22 }$(mV) & 2.4 & - & - & -\\
       $\tau_{22}$(\SI{}{\micro s}) & 0.244 & - & - & -\\
       $T_{2}$(\SI{}{\micro s}) & 1.286 & - & - & -\\
       $\phi_{2}$ & -1.683 & - & - & -\\
       \hline
       $A_{13}$(mV) & 14.6 &  & -15.3 & 1.3\\
       $\tau_{13}$(\SI{}{\micro s}) & 0.017 & & 3.51 & 3.42\\
       \hline
       $A_{14}$(mV) & - & - & 1.4 & -\\
       $\tau_{14}$(\SI{}{\micro s}) & - &- & 14.9 & -\\
         \hline
         \hline
    Device B & CPLR & QB2 & CPLR $\rightarrow$ QB2 & QB2 $\rightarrow$ CPLR\\
        \hline
      $A_{11} $(mV) & -17.65  & -3.50 & 158.18 & - \\
       $\tau_{11}$(\SI{}{\micro s}) & 179.8 & 194.1 & 198.13 & - \\
       \hline
       $A_{12}$(mV) & -7.40 & -0.967 & 19.87  & - \\
       $\tau_{12}$(\SI{}{\micro s}) & 59.3 & 73.8 & 57.2 & - \\
       \hline
       $A_{13}$(mV) & -2.61 & -10.1 & 11.0 & - \\
       $\tau_{13}$(\SI{}{\micro s}) & 3.58 &0.291 & 5.83 & \\
       \hline
       $A_{14}$(mV) & -6.51 & - & - & -\\
       $\tau_{14}$(\SI{}{\micro s}) & 0.40 & - & - & -\\
       \hline
       $A_{15}$(mV) & -15.3 & - & - & -\\
       $\tau_{15}$(\SI{}{\micro s}) & 54.4 &- & - & - \\
       \hline

         \hline
         \hline

        \end{tabular}
    \caption{Flux transient calibration result for the coupler and tunable transmon (qubit 2) in device A and device B. In device B, we used purely exponential decay without the oscillating term to model the flux transient.}
    \label{tab:flux_transient_calibration}
\end{table}

\section{CZ Gate Calibration and Benchmarking}
\label{app:2qb_czgate_calibration}

Calibrating the CZ gate in a TFT system involves finding the optimal pulse parameters to achieve an accurate conditional phase and minimal leakage. Here, we use the calibration process to construct a CZ gate based on a Slepian-shaped flux pulse \cite{Martinis2014_Slepian} as an example, which is known to effectively reduce the non-adiabatic transition. We optimize two Slepian parameters: Slepian bandwidth and the final frequency. The Slepian bandwidth determines the spectral content of the pulse. In this research, it is controlled by sweeping the standardized half-bandwidth parameter NW from the scipy.signal.dpss package. The final frequency refers to the detuning between the two relevant energy levels at the point of maximum flux amplitude. That is, the final frequency is zero when the peak amplitude of the CZ gate flux pulse brings the system precisely to the avoided crossing.

\begin{figure*}[t]
    \centering
    \includegraphics[width=0.98 \textwidth]{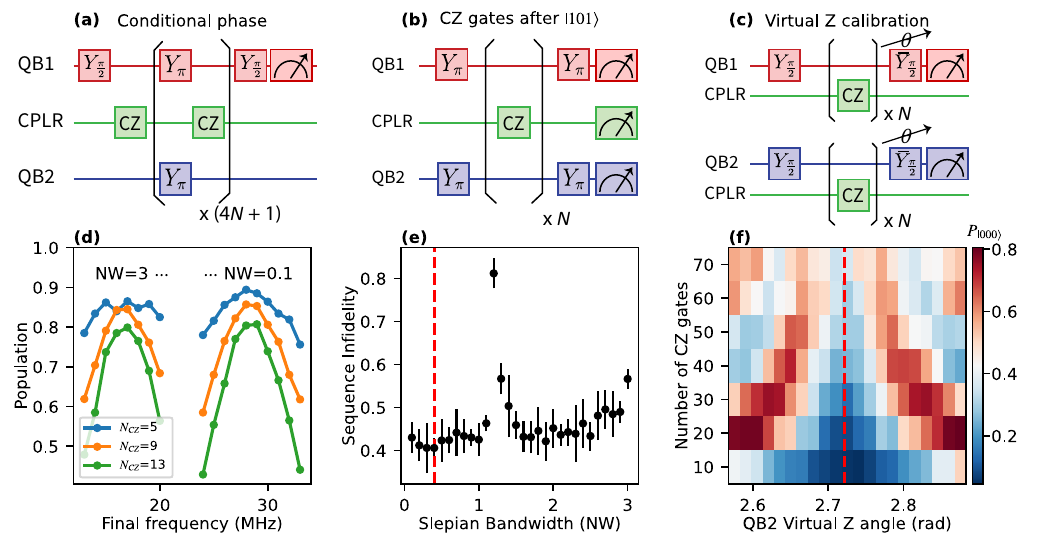}
    \caption{(a, d) Conditional phase calibration. For a selected range of standardized half bandwidth (NW), the final frequency was calibrated to achieve 180-degree conditional phase. A JAZZ pulse sequence is used to amplify the phase error for a more precise calibration. (b, e) Slepian bandwidth optimization. The sequence fidelity is measured for an initial state $|101 \rangle$ after a sequence of $N=40$ CZ gates. The Slepian bandwidth that minimizes this infidelity is selelcted as the optimal value. (c, f) Virtual Z-angle calibration for qubit 1 and qubit 2. We create a two-dimensional map of the population after $Y_{\pi/2}$, $Y_{-\pi/2} = \bar{Y}_{\pi/2}$ pulses with $N$ CZ gates in between. The virtual Z angle is determined by finding the phase (depicted by the arrow over $\theta$) of the second $\bar{Y}_{\pi/2}$ pulse that minimizes the measured population.}
    \label{fig:2qb_czgate_calibration_with_pulse}
\end{figure*}

As the first step of calibration, for each Slepian bandwidth, we adjust the final frequency to produce a 180-degree conditional phase. To amplify conditional phase error and improve calibration accuracy, we use JAZZ (Joint Amplification of $ZZ$ interaction) pulse sequence \cite{RuiLi2024DTSCZGate} with a varying number of CZ gates between the two $Y_{\pi/2}$ pulses [\fref{fig:2qb_czgate_calibration_with_pulse}(a),(b)]. 

Next, we find the optimal Slepian bandwidth by preparing the initial state $|101 \rangle$ and measuring the fidelity of $N$ CZ gates. The Slepian bandwidth that maximizes the fidelity is chosen as the optimum value [\fref{fig:2qb_czgate_calibration_with_pulse}(c),(d)]. The fidelity is determined by measuring the population that returns to the ground state after the sequence. Finally, with the optimized Slepian parameters, we measure the single-qubit phase accumulation during each CZ gate and compensate for it by applying virtual Z-gates to each qubit. [\fref{fig:2qb_czgate_calibration_with_pulse}(e),(f)].

After calibration, we assess the CZ gate using two-qubit interleaved randomized benchmarking protocol. The two-qubit Clifford gates are randomly sampled from 11520 gates, which are generated by \{$I, \pm X, \pm Y, \pm X_{\pi/2}, \pm Y_{\pi/2}$, CZ\}.

\begin{figure}[t]
    \centering
    \includegraphics[width=0.98\linewidth]{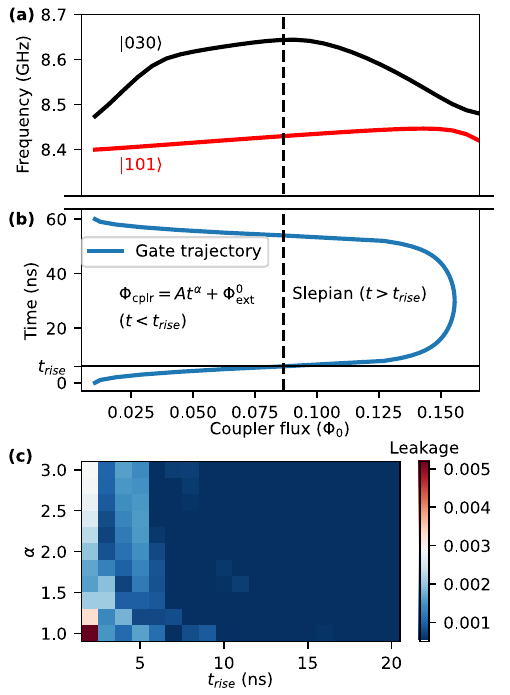}
    \caption{(a) TFT system simplified as a two-level system. (b) TFT gate trajectory as a function of time and coupler flux. (c) Numerican simulation of leakage to $|030\rangle$ during $t< t_{\textrm{rise}}$. The time-dependent numerical simulation is done using QuTip package \cite{lambert2025qutip5}.} 
    \label{fig:TFT_gate_trajectory}
\end{figure}

We first prepare a random two-qubit Clifford sequence (the reference sequence) that ends with a recovery gate that makes the overall gate sequence equivalent to an identity gate. The reference sequences, with different numbers of Cliffords, are applied to the initial state $|000 \rangle$. The infidelity of the reference sequence, $r_{\textrm{ref}}$, is calculated by fitting the ground state population with $A p_{\textrm{ref}}^{m} + B$, where $m$ is the number of Cliffords in the sequence and $r_{\textrm{ref}} = 1 - (1-1/d) p_{\textrm{ref}}$. Here, $d=4$ is the number of basis in the two-qubit Hilbert space.

Next, we interleaved a CZ gate between each two-qubit Clifford of the reference sequence and similarly calculated the sequence infidelity, $r_{\textrm{int}}$. Repeating this process over multiple randomized sequences of two-qubit Cliffords yields the average fidelity. We estimate the CZ gate error by comparing the average fidelity of the reference sequence and the interleaved sequence \cite{RuiLi2024DTSCZGate}:
\begin{align}
    r_{CZ} =\frac{d - 1}{d} \left( 1 - \frac{p_{\textrm{int}}}{p_{\textrm{ref}}} \right).
\end{align}

\section{CZ Gate Flux Pulse Construction}
\label{app:flux_pulse_optimization}

In this work, the Slepian flux pulse was designed based on a two-level system consisting of $|101\rangle$ and $|030\rangle$. However, the energy difference between these two states varies nonmonotonically as the external flux is swept. At the idling point, which is a starting point of the CZ gate, there is \SI{75}{MHz} frequency difference between the two states. As we increase the coupler flux, the frequency difference grows to a maximum of \SI{220}{MHz} at $\Phi_{\textrm{ext}} = 0.090 \Phi_{0}$. Finally, as the pulse approaches the avoided crossing between the two states ($\Phi_{\textrm{ext}} \approx 0.16 \Phi_{0}$), the detuning decreases to approximately \SI{50}{MHz}.

Thus, we construct flux pulses in two separate segments and then combine them. The first segment ($t \in[0, t_{\mathrm{rise}}]$) spans from the idling point to the flux bias point of maximum frequency difference between $|101\rangle$ and $|030 \rangle$ states ($\Phi_{\textrm{ext}} = 0.090 \Phi_{0}$). The second segment ($t\in [t_{\mathrm{rise}}, t_{\mathrm{CZ}}/2]$)continues from $\Phi_{\textrm{ext}} = 0.090 \Phi_{0}$ to the turnaround flux bias ($\Phi_{\textrm{ext}}\approx 0.16 \Phi_{0}$). 

For the first segment, we define the pulse shape as a polynomial in time: $\Phi_{\textrm{ext}}(t) = At^{\alpha}+\Phi_{\textrm{ext}}^{0}$. To ensure adiabaticity, we performed a numerical simulation to find the values of $t_{\mathrm{rise}},\alpha$ that minimize leakage during the rising time. Although the simulation suggests that leakage is minimized at $t_{\textrm{rise}}>$\SI{10}{ns}, we experimentally found that the maximum fidelity was obtained with a much shorter rising time of approximately $t_{\mathrm{rise}} \approx $\SI{2}{ns} and $\alpha=2$. We attributed this to the fact that for longer pulses, the reduction in leakage error is outweighed by the increase in decoherence. For a given $\alpha, t_{\mathrm{rise}}$, $A$ is determined such that $\Phi_{\mathrm{ext}}(t=t_{\mathrm{rise}})=0.090 \Phi_{0}$. Next, at $t > t_{\mathrm{rise}}$ we construct a Slepian-shaped flux pulse using the optimized final frequency and Slepian bandwidth.

\section{Origin of $ZZ$ Interaction in a TFT system}
\label{app:zz_analysis}
We can quantitatively analyze the $ZZ$ interaction of a TFT system by treating the interaction Hamiltonian, $\hat{H}_{\textrm{int}} = J_{1c} \hat{n}_{1} \hat{n}_{c} + J_{2c} \hat{n}_{c} \hat{n}_{2} + J_{12} \hat{n}_{1} \hat{n}_{2}$, as a perturbation. Perturbative calculation provides insight into the higher-order interactions responsible for the cancellation and enhancement of the overall $ZZ$ interaction.

To illustrate, the second-order perturbative expansion of the energy level of state $| k\rangle$ can be written as follows:
\begin{align}
    \delta E_{k}^{(2)} = \sum_{l \neq k} \frac{|\langle l | \hat{H}_{\textrm{int}} |k \rangle |^{2}}{E_{k}^{0} - E_{l}^{0}},
\end{align}
where $E_{k}^{0}$ is the bare energy of the kth level in a TFT system, which assumes zero interaction energy. Therefore, the $ZZ$ interaction from the second-order perturbation can be written as below:

\begin{align}
    \hbar \zeta_{ZZ}^{(2)} = \delta E_{101}^{(2)} - \delta E_{100}^{(2)} - \delta E_{001}^{(2)} + \delta E_{000}^{(2)}.
\end{align}

Similarly, we can calculate the third and fourth-order contributions, $\zeta_{ZZ}^{(3)}$ and $\zeta_{ZZ}^{(4)}$. A comparison between this perturbative estimation and a full numerical calculation of the $ZZ$ interaction is shown in \fref{fig:zz theory vs perturb}. The figure shows that the second-order and third-order perturbation theory fails to estimate the numerical result while including the fourth-order perturbation calculation reproduces the correct profile. This discrepancy is because the $ZZ$ interaction in a TFT system is strongly influenced by states in the multiple-excitation manifold states $\{|101\rangle$, $|020\rangle$, $|030\rangle \}$, which cannot show up in the expansions smaller than fourth order. For instance, at zero coupler flux, the dominant contribution of the fourth-order contribution, $\zeta_{ZZ}^{(4)}$, satisfies the following expression:

\begin{figure}
    \centering
    \includegraphics[width = 0.48 \textwidth]{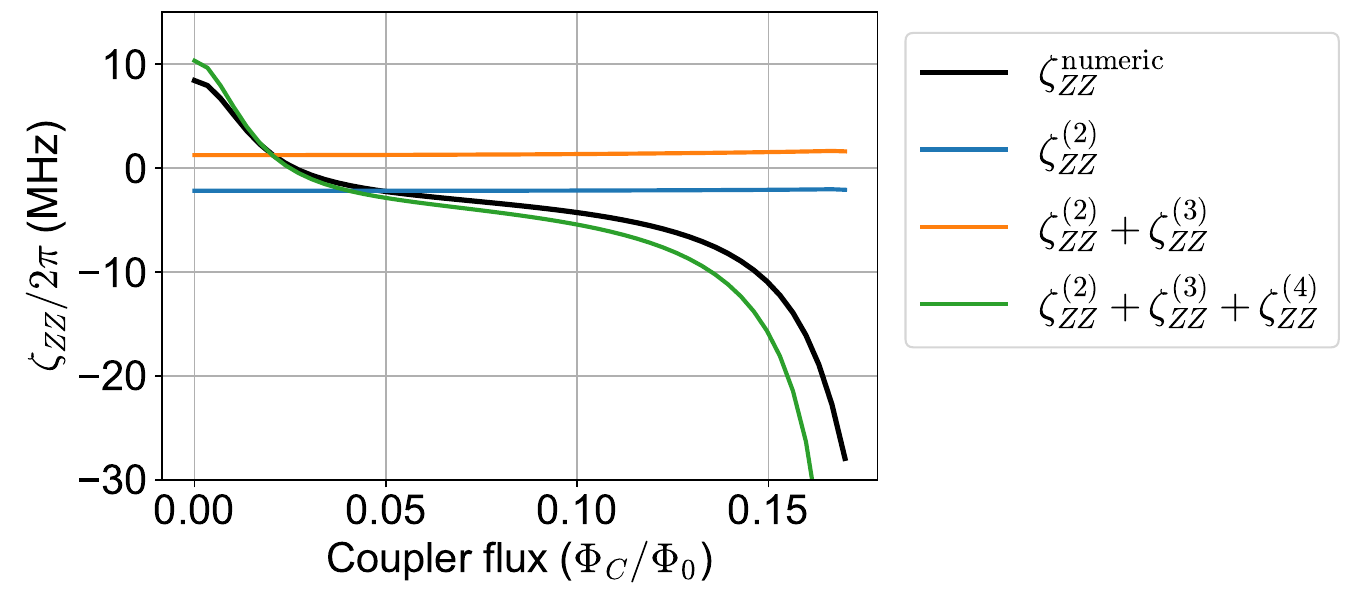}
    \caption{Calculating ${ZZ}$ interaction in a TFT system: numerical diagnoalization (solid black line) versus perturbation theory. The parameters are identical to \fref{fig:transition_zz_vs_cplrflux}. The plot shows that the second-order and third-order perturbation theory ($\zeta_{ZZ}^{(2)}, \zeta_{ZZ}^{(3)}$) fail to capture the $ZZ$ interaction profile, while including $\zeta_{ZZ}^{(4)}$ reproduces the general trend.}
    \label{fig:zz theory vs perturb}
\end{figure}

\begingroup\makeatletter\def\f@size{8}\check@mathfonts
\def\maketag@@@#1{\hbox{\m@th\small\normalfont#1}}%

\begin{align}
\begin{split}
    &\zeta_{ZZ}^{(4)} = \cdots + \\
    &\frac{|\langle 101| \hat{H}_{\textrm{int}}|011\rangle \langle 011| \hat{H}_{\textrm{int}} | 020\rangle \langle 020| \hat{H}_{\textrm{int}} | 011\rangle \langle 011| \hat{H}_{\textrm{int}} | 101\rangle|^{2} }{(E_{101}^{0} - E_{011}^{0})(E_{101}^{0} - E_{020}^{0})(E_{101}^{0} - E_{011}^{0})}. 
\end{split}
\end{align}
\endgroup

\noindent The term above describes the higher-order interaction between $|101\rangle$ and $|020\rangle$ mediated by $|011\rangle$. This term has a positive sign, meaning that the energy of $|101\rangle$ state is effectively "pushed up" by this interaction. Table~\ref{tab:tft_zz_contribution_zeroflux} and~\ref{tab:tft_zz_contribution_atop} list the seven fourth-order perturbative terms that contribute most significantly to the $ZZ$ interaction at two key bias points: zero external flux and near the avoided crossing between $|101\rangle$ and $|030\rangle$. 

\begin{center}
\begin{table}[h]
    \begin{tabular}{>{\centering}m{6.0cm} >{\centering\arraybackslash}m{2.0cm}}
        \hline \hline
        states involved & $\Delta \zeta_{ZZ}/2 \pi$ (MHz) \\ 
        \hline
       $|101\rangle \leftrightarrow |011\rangle \leftrightarrow |020\rangle \leftrightarrow |011\rangle \leftrightarrow |101\rangle $ & 4.113 \\ 
       $|101\rangle \leftrightarrow |011 \rangle \leftrightarrow |020\rangle \leftrightarrow |110 \rangle \leftrightarrow |101\rangle $ & 2.118 \\
       $|101\rangle \leftrightarrow |110 \rangle \leftrightarrow |020\rangle \leftrightarrow |011 \rangle \leftrightarrow |101\rangle $ & 2.118 \\
       $|101\rangle \leftrightarrow |011 \rangle \leftrightarrow |002\rangle \leftrightarrow |011 \rangle \leftrightarrow |101\rangle $ & 1.996 \\
       $|100\rangle \leftrightarrow |010 \rangle \leftrightarrow |001\rangle 
       \leftrightarrow |010 \rangle \leftrightarrow |101\rangle $ & -1.397 \\
       $|101\rangle \leftrightarrow |110 \rangle \leftrightarrow |200\rangle 
       \leftrightarrow |110 \rangle \leftrightarrow |101\rangle $ & -1.235 \\
       $|101\rangle \leftrightarrow |110 \rangle \leftrightarrow |020\rangle \leftrightarrow |110 \rangle \leftrightarrow |101\rangle $ & 1.090 \\
        \hline \hline
    \end{tabular}
    \caption{The seven-largest terms of the 4th-order perturbative expansion of $ZZ$ interaction. The expansion is done at zero external flux. The first three largest $ZZ$ interaction contributions comes from the higher-order interaction between $|101\rangle$ and $|020\rangle$.}
    \label{tab:tft_zz_contribution_zeroflux}
    \end{table}
\end{center}

\begin{center}
\begin{table}[h]
    \begin{tabular}{>{\centering}m{6.0cm} >{\centering\arraybackslash}m{2.0cm}}
        \hline \hline
        states involved &  $\Delta \zeta_{ZZ}/2 \pi$ (MHz) \\ 
        \hline
       $|101\rangle \leftrightarrow |011\rangle \leftrightarrow |030\rangle \leftrightarrow |011\rangle \leftrightarrow |101\rangle $ & -36.18 \\ 
       $|101\rangle \leftrightarrow |011 \rangle \leftrightarrow |030\rangle \leftrightarrow |110 \rangle \leftrightarrow |101\rangle $ & -17.22 \\
       $|101\rangle \leftrightarrow |110 \rangle \leftrightarrow |030\rangle \leftrightarrow |011 \rangle \leftrightarrow |101\rangle $ & -17.22 \\
       $|101\rangle \leftrightarrow |110 \rangle \leftrightarrow |030\rangle \leftrightarrow |110 \rangle \leftrightarrow |101\rangle $ & -8.198 \\
       $|101\rangle \leftrightarrow |011 \rangle \leftrightarrow |002\rangle \leftrightarrow |011 \rangle \leftrightarrow |101\rangle $ & 2.510 \\
       $|100\rangle \leftrightarrow |010 \rangle \leftrightarrow |001 \rangle \leftrightarrow |010 \rangle \leftrightarrow |101\rangle $ & -1.757 \\
       $|101\rangle \leftrightarrow |110 \rangle \leftrightarrow |200 \rangle \leftrightarrow |110 \rangle \leftrightarrow |101\rangle $ & -1.327 \\
        \hline \hline
    \end{tabular}
    \caption{The seven-largest terms of the 4th-order perturbative expansion of $ZZ$ interaction. The expansion is done at the external flux near avoided crossing between $|101\rangle$ and $|030\rangle$ ($\Phi_{C}/\Phi_{0} = 0.17$), where $ZZ$ interaction strength is -27.9 MHz}
    \label{tab:tft_zz_contribution_atop}
    \end{table}
\end{center}

\newpage 

\section{$ZZ$ Interaction in Straddling Regime}
\label{app:TFT_$ZZ$_straddling regime}

\begin{figure}[H]
    \centering
    \includegraphics[width=0.98\linewidth]{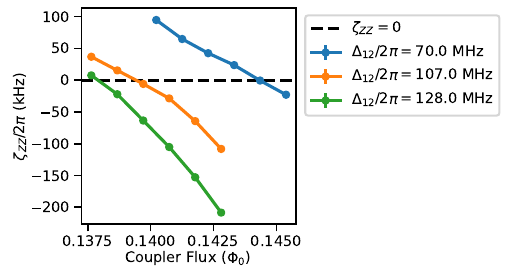}
    \caption{$ZZ$ interaction vs coupler flux in device A when the system is in the straddling regime. We set the qubit-qubit detuning of \SI{70}{MHz}, \SI{107}{MHz}, \SI{128}{MHz}, and confirmed points where $ZZ$ interaction changes sign for each case.}
    \label{fig:zz_at_straddling_regime}
\end{figure}

In the main text, we identified a zero-$ZZ$ point when the system is outside of the straddling regime. However, a TFT system is also capable of completely canceling the $ZZ$ interaction even when the detuning is smaller than the qubit anharmonicities (i.e., straddling regime). To demonstrate this, we chose three different qubit-qubit detuning values for device A by tuning the qubit 2 frequency close to the qubit 1 frequency so the system is in the straddling regime. We experimentally verified the existence of an idling point in each case [\fref{fig:zz_at_straddling_regime}]. 

\section{Summary of TFT Device B}
\label{app:tft1_summary}

\begin{figure*}[t]
    \centering
    \includegraphics[width=0.98 \linewidth]{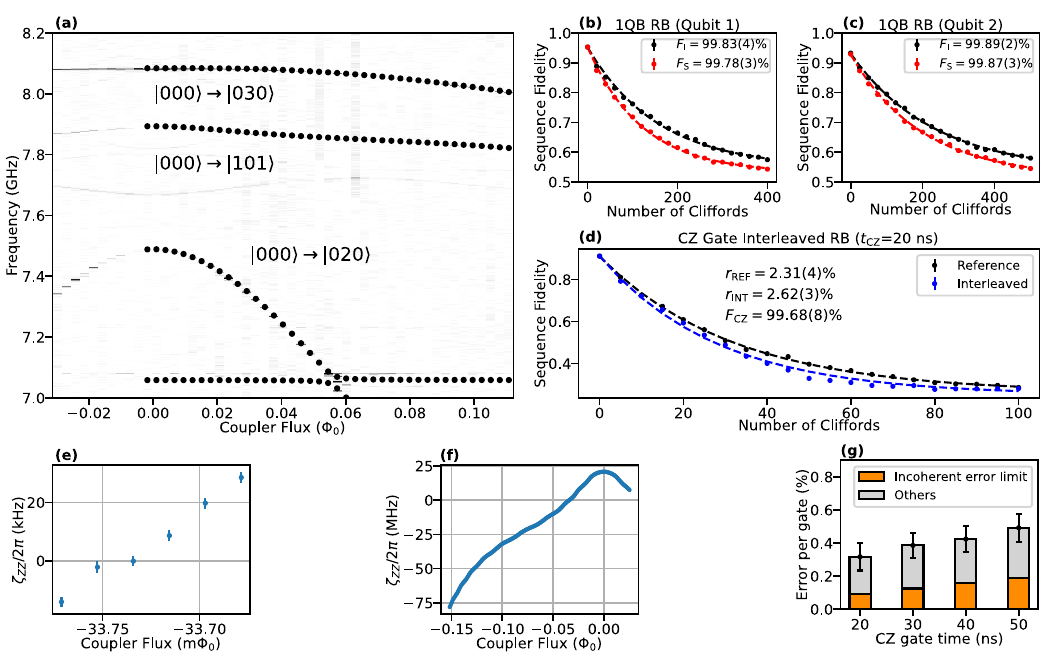}
    \caption{Summary of measurement results for device B. (a) Qubit spectrum of device B and its corresponding numerical fit. The plot shows the multi-photon excitation manifold, including $|020\rangle, |101\rangle,|030\rangle$. (b - c) Single-qubit randomized benchmarking of qubit 1 (b) and qubit 2 (c). (d) Interleaved randomized benchmarking of the CZ gate, showing gate fidelity of 99.68(8)\% for \SI{20}{ns} gate time. Single-qubit gates were implemented using a cosine-shaped pulse of \SI{50}{ns} duration. (e - f) $ZZ$ interaction measurement results near zero-$ZZ$ point (e) and over a wider range of coupler flux range (f). We measured zero $ZZ$ interaction at $\Phi_{c} \approx -33.73 \mathrm{m \Phi_{0}}$. (g) CZ gate error for four different gate times. The grey area ``Others" may include leakage error or coherent errors such as miscalibration, remaining flux transient, and so on. The results of the single-qubit gate and the two-qubit gate randomized benchmarking were averaged over 20 random two-qubit Clifford sequences. The error bars of the plots specify the standard error of the means, similar to \fref{fig:cz_rb} from the main text.}   
    \label{fig:tft1_deviceB_summary}
\end{figure*}

In this section, we summarize the measurement results of device B. As shown in Table~\ref{tab:tft_param}, device B has an interaction strength nearly twice as large as that of device A, enabling stronger hybridization between the qubit and the coupler states. Consequently, we were able to measure a $ZZ$ interaction of \SI{-75}{MHz} at the coupler flux of $-0.15 \Phi_{0}$, which is almost four times higher than the $ZZ$ interaction measured at device A [\fref{fig:tft1_deviceB_summary}(f)]. We could also find the coupler flux bias that completely cancels the $ZZ$ interaction at $\Phi_{c} \approx -33.73\mathrm{m \Phi_{0}}$ [\fref{fig:tft1_deviceB_summary}(e)]. 

After confirming the zero-$ZZ$ point of device B, we calibrated and benchmarked the single-qubit and two-qubit gates. For qubit 1, we obtained a single-qubit gate fidelity of 99.83(4)\% from individual, and 99.78(3)\% from simultaneous randomized benchmarking [\fref{fig:tft1_deviceB_summary}(b)]. For qubit 2, we measured fidelities of 99.89(2)\% and 99.87(3)\% for individual and simultaneous randomized benchmarking, respectively [\fref{fig:tft1_deviceB_summary}(c)]. The small fidelity difference between the individual and simultaneous benchmarking results is consistent with the suppressed $ZZ$ interaction. Next, we benchmarked the two-qubit CZ gate using the interleaved randomized benchmarking protocol. At a \SI{20}{ns} gate time, we obtained an average gate fidelity of 99.68(8)\% [\fref{fig:tft1_deviceB_summary}(d)] with a cosine-shaped flux pulse. Additionally, we measured average CZ gate errors less than 0.5\% for gate times ranging from \SI{20}{ns} to \SI{50}{ns} [\fref{fig:tft1_deviceB_summary}(g)].

Unfortunately, device B was damaged before we could apply pulse optimization and complete the gate error analysis. The device could not be repaired, so we focused the remainder of our study on device A.

\clearpage

\end{document}